\documentclass[9pt,twocolumn,twoside]{optica}
\setboolean{shortarticle}{false}
\setboolean{minireview}{true}

\usepackage{caption,upgreek,verbatim,braket,cancel,dblfloatfix,hyperref, nicefrac}

\title{Cavity quantum electrodynamics with color centers in diamond}

\author[1, $\dagger$]{Erika Janitz}
\author[2, $\dagger$]{Mihir K. Bhaskar}
\author[1, *]{Lilian Childress}

\affil[1]{Department of Physics, McGill University, 3600 Rue University, Montreal QC, H3A 2T8, Canada}
\affil[2]{Department of Physics, Harvard University, 17 Oxford St, Cambridge MA, 02138, USA}
\affil[$\dagger$]{These authors contributed equally}
\affil[*]{Corresponding author: lilian.childress@mcgill.ca}

\begin{abstract}
   Coherent interfaces between optical photons and long-lived matter qubits form a key resource for a broad range of quantum technologies. Cavity quantum electrodynamics (cQED) offers a route to achieve such an interface by enhancing interactions between cavity-confined photons and individual emitters. Over the last two decades, a promising new class of emitters based on defect centers in diamond have emerged, combining long spin coherence times with atom-like optical transitions. 
   More recently, advances in optical resonator technologies have made it feasible to realize cQED in diamond. This article reviews progress towards coupling color centers in diamond to optical resonators, focusing on approaches compatible with quantum networks.  We consider the challenges for cQED with solid-state emitters and introduce the relevant properties of diamond defect centers before examining two qualitatively different resonator designs: micron-scale Fabry-Perot cavities and diamond nanophotonic cavities. For each approach, we examine the underlying theory and fabrication, discuss strengths and outstanding challenges, and highlight state-of-the-art experiments.
\end{abstract}

\begin{document}

\maketitle

\section{Introduction}

The last two decades have seen an explosive growth in quantum technologies, with remarkable advances in cryptography~\cite{gisin_quantum_2002, scarani_security_2009, lo_secure_2014, pirandola_advances_2019}, computing~\cite{ladd_quantum_2010}, and metrology~\cite{degen_quantum_2017}.
Motivated by the success of the classical internet and the promise of quantum-secured communication, considerable interest has arisen in developing so-called quantum networks~\cite{kimble_quantum_2008, wehner_quantum_2018}. Such networks comprise nodes capable of storing and processing quantum information, connected by quantum-coherent photonic channels, and could enable distributed quantum computing
~\cite{monroe_large-scale_2014, nemoto_photonic_2014, fitzsimons_private_2017}, quantum-enhanced clock synchronization~\cite{jozsa_quantum_2000,  giovannetti_quantum-enhanced_2001, komar_quantum_2014} 
and interferometry~\cite{gottesman_longer-baseline_2012, khabiboulline_optical_2019}, as well as long-distance quantum communication~\cite{briegel_quantum_1998, munro_inside_2015, muralidharan_optimal_2016}. 

\begin{figure}
	\begin{center}
		\includegraphics[scale=1]{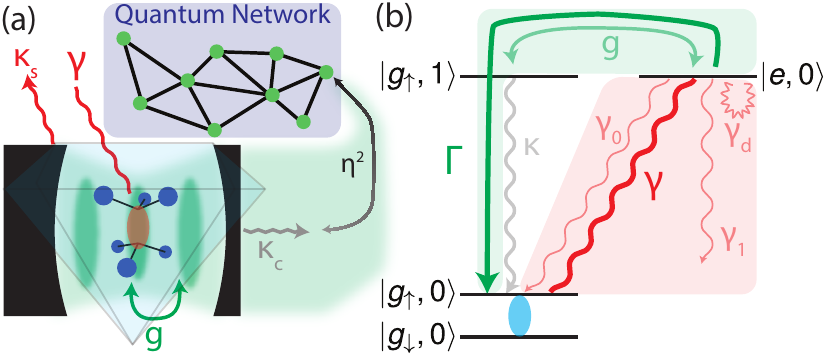}
		\caption{\label{fig:levels}
			(a) Schematic of diamond cQED system. Cavity photons are coherently coupled to the color center at rate $g$ and can scatter out of the cavity at rate $\kappa_s$, or into a collected mode at rate $\kappa_c$. Optical decoherence of the color center occurs at rate $\gamma$. Indistinguishable photons in the collected mode are matched into a single-mode fiber with efficiency $\eta^2$.
			(b) Detailed diagram for one instance of relevant cQED energy levels, comparing coherent coupling rates (green) to decoherence rates (gray, red). Quantum information (blue oval) is stored in the color center spin states $\ket{g_\downarrow, 0}, \ket{g_\uparrow, 0}$, where $g (e)$ indicates the ground (excited) state, $\uparrow, \downarrow$ indicate spin, and the final integer indicates cavity photon number.  $\ket{e,0}$ couples spin-selectively to $\ket{g_\uparrow, 1}$ at rate $g$, leading to an effective cavity-enhanced emission rate $\Gamma$. The cavity dissipates at overall decay rate $\kappa = \kappa_s + \kappa_c$ and the emitter dephases at overall rate $\gamma$, which comprises desired emission ($\gamma_0$), undesired decay ($\gamma_1$), and pure dephasing ($\gamma_d$).
		}
	\end{center}
\end{figure}

Optical photons play a critical role in quantum networks because they interact very weakly with surrounding media, making them the ideal carrier of quantum information over long distances. 
However, these weak interactions pose a significant impediment for engineering gates between photons, and make it difficult to interface photons with information storage and processing nodes.
Cavity quantum electrodynamics (cQED) offers a paradigm to overcome these challenges: by confining photons inside a high quality factor optical resonator, vastly enhanced interactions with material systems can be obtained~\cite{goy_observation_1983, kimble_strong_1998, mabuchi_cavity_2002, vahala_optical_2003,  lodahl_interfacing_2015, reiserer_cavity-based_2015}, 
enabling a broad range of relevant quantum information tasks~\cite{pellizzari_decoherence_1995, cirac_quantum_1997, northup_quantum_2014}. For example, cavity-coupled emitters can be used as bright sources of the indistinguishable photons~\cite{mckeever_deterministic_2004, senellart_high-performance_2017} needed for photonic quantum computation~\cite{knill_scheme_2001, kok_linear_2007}, or even employed to mediate interactions between photons~\cite{dayan_photon_2008}. The most exciting opportunities emerge when the material system possesses long-lived internal degrees of freedom, such as electron or nuclear spin sublevels, in addition to its optical transitions [an example is shown in Fig.~\ref{fig:levels}(b)]. Such long-lived states can be used to realize a quantum memory for light~\cite{specht_single-atom_2011, clausen_quantum_2011, chen_coherent_2013, seri_quantum_2017}, spin-photon entanglement~\cite{blinov_observation_2004, matsukevich_entanglement_2005, volz_observation_2006, togan_quantum_2010, gao_observation_2012, de_greve_quantum-dot_2012}, spin-photon switches~\cite{tiecke_nanophotonic_2014, sun_quantum_2016}, or quantum gates between asynchronous photons~\cite{duan_scalable_2004, beck_large_2016, hacker_photonphoton_2016}. 
While some of these capabilities can be realized without cavities (e.g. using optically dense ensembles~\cite{sangouard_quantum_2011, de_riedmatten_quantum_2015}), 
cQED can enhance the efficiency of probabilistic protocols, and even enable near-deterministic interactions between single photons and individual quantum bits.



The likely impact of cQED systems on quantum networks is exemplified by their potential role in quantum repeaters, which compensate for photon loss to enable transmission of quantum states or secret keys over long distances~\cite{briegel_quantum_1998, munro_inside_2015, muralidharan_optimal_2016}.  The simplest repeater schemes distribute entanglement over long distances via a chain of spins linked by photonic channels~\cite{briegel_quantum_1998}, generating a resource that can be purified~\cite{bennett_purification_1996, reichle_experimental_2006, kalb_entanglement_2017} and used to teleport quantum information~\cite{bennett_teleporting_1993, bouwmeester_experimental_1997} or generate a secret key~\cite{ekert_quantum_1991}. Such schemes often employ heralded entanglement generation ~\cite{cabrillo_creation_1999, barrett_efficient_2005, chou_functional_2007, moehring_entanglement_2007, hofmann_heralded_2012,  bernien_heralded_2013, delteil_generation_2016, stockill_phase-tuned_2017}, which exploits entanglement between spins and indistinguishable outgoing photons, such that a detected photon could have originated from either of two distant spins. Conditioned on a detection event, the spins are projected onto an entangled state, and photon loss errors result only in reduced success probability.
Cavity coupling can vastly increase the efficiency of these protocols by increasing photon emission rates into a well-defined cavity-coupled mode~\cite{purcell_spontaneous_1946, goy_observation_1983}; for strongly-coupled cQED systems, even deterministic remote entanglement generation can be possible~\cite{cirac_quantum_1997, axline_-demand_2018}. 
In the longer term, one-way repeater schemes promise much faster communication by transmitting entangled multi-photon states that encode quantum information in a photon-loss-tolerant structure~\cite{varnava_loss_2006, fowler_surface_2010}, potentially incorporating error correction
~\cite{muralidharan_ultrafast_2014}, and without the need for long-term quantum memory~\cite{munro_quantum_2012, azuma_all-photonic_2015, pant_rate-distance_2017}. 
In this context, cavity-coupled emitters play an essential role: by maintaining a coherent quantum memory during repeated interactions with light, an emitter can generate entanglement between sequentially emitted photons~\cite{schon_sequential_2005, lindner_proposal_2009, schwartz_deterministic_2016, buterakos_deterministic_2017} and be used for efficient re-encoding at one-way repeater stations~\cite{ borregaard_one-way_2020}. 
While these are just a subset of quantum repeater and networking functionalities being actively pursued, they illustrate the key role played by cavity-coupled spin-photon interfaces in near-term quantum information applications~\cite{borregaard_quantum_2019}. 

A wide variety of physical platforms for realizing cQED are currently being explored, aimed at achieving interaction rates between the quantum emitter and cavity mode that exceed relevant losses. 
Cavity-enhanced interfaces between optical photons and individual quantum emitters began with neutral atoms~\cite{reiserer_cavity-based_2015} and quickly expanded to other systems including quantum dots~\cite{lodahl_quantum-dot_2017}, molecules~\cite{wang_turning_2019}, and trapped ions~\cite{takahashi_strong_2020,kobel_deterministic_2020}. In particular, there has been a recent interest in cavity coupling to atomic-like solid-state systems~\cite{awschalom_quantum_2018, atature_material_2018}, most notably defects in diamond~\cite{schroder_quantum_2016, johnson_diamond_2017, bradac_quantum_2019}, but also rare-earth ions~\cite{zhong_optically_2018, dibos_atomic_2018}, defects in silicon carbide~\cite{crook_purcell_2020}, 
and potentially others~\cite{bassett_quantum_2019}. These defect-based systems aim to combine the atomic-like advantages of a predictable structure, optical selection rules and long spin coherence times with a robust solid-state platform compatible with integration into micro- or nano-scale cavities. The various host materials and defect structures carry different strengths, as discussed in recent reviews~\cite{atature_material_2018, awschalom_quantum_2018}. Here, we focus on the most well-studied platform for defect-based, optically-active spin qubits: diamond. 

Indeed, diamond's large bandgap and nearly nuclear-spin-free lattice make it an appealing host for such defects \cite{zaitsev_optical_2001}. 
Of these, the best known is the nitrogen-vacancy (NV) center, which exhibits long spin coherence times, access to nearby nuclear spins for ancilla qubits~\cite{bradley_ten-qubit_2019}, and well-understood spin-selective optical transitions~\cite{maze_properties_2011, doherty_negatively_2011}. Thanks to these properties, cavity-coupled NVs form the basis for many quantum information proposals \cite{childress_fault-tolerant_2006, nemoto_photonic_2014, lo_piparo_memory-assisted_2017, rozpedek_near-term_2019}.
Moreover, even without cavity coupling, NVs have already been used to achieve landmark proof-of-principle quantum network experiments demonstrating loophole-free Bell inequality violation~\cite{hensen_loophole-free_2015}, unconditional teleportation~\cite{pfaff_unconditional_2014}, entanglement distillation~\cite{kalb_entanglement_2017}, and entanglement distribution faster than entanglement loss~\cite{humphreys_deterministic_2018}, all of which would exhibit vastly improved efficiency using cavity coupling. 
At the same time, novel defects such as the silicon vacancy (SiV) have been found to exhibit superior optical properties~\cite{bradac_quantum_2019} and the potential for long spin coherence times~\cite{sukachev_silicon-vacancy_2017, rose_observation_2018, metsch_initialization_2019}. 
These advances have helped to motivate development of cQED systems in diamond.

    This article reviews recent progress in coupling individual diamond defects to optical resonators, focusing on cQED platforms with a clear potential for quantum applications. Over the last decade diamond defects have been coupled to a wide range of photonic and plasmonic  systems, and we refer the reader to excellent reviews~\cite{schroder_quantum_2016, johnson_diamond_2017, huck_coupling_2016} for an overview. Here, we focus on two approaches with the strongest near-term promise for quantum networking applications.  One avenue employs external Fabry-Perot microcavities~\cite{trupke_microfabricated_2005, hunger_fiber_2010}  surrounding the defect center. While early work with defects in nanodiamond laid crucial groundwork~\cite{albrecht_coupling_2013,albrecht_narrow-band_2014,johnson_tunable_2015,kaupp_purcell-enhanced_2016,benedikter_cavity-enhanced_2017}, the chief advantage of Fabry-Perot micro-cavities for quantum networking applications is that they can enclose relatively thick ($\sim \upmu$m) diamond membranes~\cite{janitz_fabry-perot_2015} where defect centers are far from charge noise at surfaces~\cite{ruf_optically_2019}. This opens the possibility to work with highly sensitive NV centers and create a cQED platform that leverages nearly two decades of advances in NV quantum science.  The second approach we consider fabricates high quality factor optical resonators from diamond itself~\cite{babinec_design_2011, faraon_resonant_2011}. While hybrid platforms combining diamond with other nano-fabricated materials hold great future promise for advanced functionality and scale-up~\cite{wolters_enhancement_2010, barclay_hybrid_2011, elshaari_hybrid_2020, kim_hybrid_2020,wan_large-scale_2020}, we focus on the recent achievements and near-term potential of all-diamond devices, which have the advantage of maximizing cavity mode overlap with defects, and have recently seen breakthroughs in fabrication techniques that have enabled major cQED milestones to be reached~\cite{nguyen_quantum_2019, nguyen_integrated_2019, bhaskar_experimental_2020}. An essential element of this recent success has been the use of novel diamond defects that are largely insensitive to proximal surface noise~\cite{bradac_quantum_2019}.  For each of these two methodologies, we review the underlying theory and fabrication considerations, highlight state-of-the-art achievements, and discuss the outstanding challenges. Indeed, recent progress in these two approaches has put the diamond defect community on the precipice of realizing key steps toward quantum networking enabled by an efficient and coherent photonic interface.

\section{Cavity QED with solid-state emitters} \label{cqed}
%
 Cavity QED enhances the coherent coupling rate between a quantum emitter and cavity-confined photons, thereby improving the efficiency with which indistinguishable photons (within the same spatial, spectral, and temporal mode) can interact with individual color center quantum memories. 
 This can approach a deterministic process when the emitter-photon interaction rate exceeds cavity losses and dephasing of the emitter optical transition. 
In this section, we examine the figures of merit for cQED with imperfect emitters, and consider different regimes for quantum applications. 

In the absence of a cavity, emitter-photon interactions scale with the spontaneous emission rate $\gamma_0$ along the desired optical transition. In the case of color centers (see Sec.~\ref{defect_section}), this is typically a transition within the zero phonon line (ZPL), 
and the 
%
%
%
 rate $\gamma_0$ is often weak compared to the overall rate of dephasing of the optical transition ($\gamma$). The total dephasing rate $\gamma$ comprises both $\gamma_0$ and  all other decay pathways ($\gamma_1)$, including nonradiative decay or emission into the phonon sideband (PSB), as well as pure dephasing ($\gamma_d$), with $\gamma = \gamma_0 +\gamma_1+\gamma_d$ [Fig.~\ref{fig:levels}(b)].

%
%
%
%
When an emitter is placed inside of an optical cavity, the local photonic density of states can be strongly enhanced at the cavity resonance frequency, enabling rapid emitter-photon interactions when the cavity is tuned to the emitter optical transition.
This enhancement can be characterized using the \emph{Purcell factor} \begin{equation} \label{purcelldef}
    \mathcal{P} = \frac{\Gamma}{\gamma_0},
\end{equation}
which compares $\Gamma$, the new rate of emission via the cavity, to $\gamma_0$, the original free-space emission rate along the relevant transition. 
Note that definitions of the Purcell factor vary, especially for non-ideal emitters, and we have chosen this definition to clearly differentiate the regimes of cQED with broadened emitters.
When the emitter is on resonance with the cavity, and the cavity decay $\kappa$ is the dominating rate, $\Gamma \approx 4 g^2/\kappa$. Here $g = \vec{\mu} \cdot \vec{\mathcal{E}_0}/\hbar$ is the rate of interaction between the dipole moment $\vec{\mu}$ of the optical transition of interest and the cavity mode vacuum field $\vec{\mathcal{E}_0}$ at the emitter location. 
Notably, since $\gamma_0\propto \mu^2$, $\mu$ drops out of $\mathcal{P}$, and for an optimally oriented and located emitter the Purcell factor is determined entirely by the properties of the cavity \cite{purcell_spontaneous_1946, fox_quantum_2006}, 
%
\begin{equation} \label{purcell}
	\mathcal{P} = \frac{3}{4\pi^2} \bigg(\frac{\lambda}{n} \bigg)^3  \bigg(\frac{Q}{V}\bigg),
\end{equation}
where $Q =\omega/\kappa$ is the quality factor of the cavity, $\omega$ is its resonance frequency, $\lambda$ is the corresponding wavelength in free space, $n$ is the index of refraction within the cavity (assumed constant), and the cavity mode volume $V$ emerges from $\mathcal{E}_0 = \sqrt{\hbar\omega/2\epsilon_0 V}$.
The scaling of $\mathcal{P} \propto Q/V$ naturally motivates the use of high quality-factor cavities with minimal mode volume.

$\mathcal{P}$ quantifies the enhancement of radiative emission on resonance with the cavity. However, for most solid-state emitters, the resonant optical emission $\gamma_0$ only accounts for a fraction of their total decay processes $\gamma_0 +\gamma_1$, implying that 
$\mathcal{P}$ does not describe the increase in the overall excited state decay rate.
Furthermore, $\mathcal{P}$ does not specify the absolute probability of coherent atom-photon interaction per attempt; this 
depends on the cooperativity $C$, where 
\begin{equation} \label{cooperativity}
	C = \frac{4g^2}{\kappa \gamma}=\mathcal{P}\bigg( \frac{\gamma_0}{\gamma} \bigg) \equiv \frac{\Gamma}{\gamma}.
\end{equation}
Here, the final equivalence (valid in the large-$\kappa$ limit),
gives a physical picture of $C$: the cooperativity compares the rate of radiation via the cavity to all emitter dephasing mechanisms.
More generally, when $C > 1$, the coherent coupling between the emitter and cavity photons is stronger than the decoherence mechanisms, leading to near-deterministic atom-photon interactions~\cite{reiserer_cavity-based_2015, borregaard_quantum_2019}.

Unlike the Purcell factor $\mathcal{P}$, the cooperativity $C$ captures the effects of the sub-optimal optical characteristics of solid-state emitters. 
In the case of an ideal, radiatively broadened two-level system, $\gamma_0/\gamma =  1$, and $C = \mathcal{P}$ exactly. However, in the case of solid-state emitters, where typically $\gamma_0/\gamma \ll 1$, 
a cQED system can have a 
Purcell factor $\mathcal{P} > 1$ but still be in the regime $C < 1$. 
There are scenarios where this intermediate regime can offer useful enhancements. For highly broadened emitters ($\gamma \gg \kappa$), such as diamond defects at room temperature, the cavity can funnel otherwise broadband emission into the relatively narrow resonator mode, creating a frequency-tunable source of 
narrow-band single photons~\cite{albrecht_coupling_2013, kaupp_scaling_2013, grange_cavity-funneled_2015, benedikter_cavity-enhanced_2017, hoy_jensen_cavity-enhanced_2020}. More typically, low temperature solid-state cQED systems operate in the $\kappa \gg \gamma$ limit, with emitters still dominated by pure dephasing $\gamma_d \gg \gamma_0, \gamma_1$. In this case, the brightness of the optical transition of interest is enhanced by $\mathcal{P}$, and when $\Gamma$ becomes larger than $\gamma_1 + \gamma_0$, 
the lifetime of the emitter begins to decrease significantly, increasing the overall rate of photon emission.
  However, as long as $C<1$, the emitter linewidth is still determined primarily by $\gamma_d$, requiring detection during a short time window $\delta t \sim 1/\gamma_d$ to render the photons indistinguishable~\cite{legero_time-resolved_2003, bernien_two-photon_2012}. Nevertheless, as the cavity coupling rate $\Gamma$ increases, so does the probability of photon emission within $\delta t$. Hence Purcell-enhanced emission, in combination with spin-selective optical transitions \cite{doherty_nitrogen-vacancy_2013, gali_ab_2019, muller_optical_2014, hepp_electronic_2014, siyushev_optical_2017, rugar_characterization_2019, trusheim_transform-limited_2020}, coherent qubit manipulation \cite{doherty_nitrogen-vacancy_2013, sukachev_silicon-vacancy_2017}, and efficient outcoupling \cite{hunger_fiber_2010, patel_efficient_2016, burek_fiber-coupled_2017}, could already greatly enhance the rate of remote entanglement generation 
~\cite{barrett_efficient_2005, bernien_heralded_2013, hensen_loophole-free_2015, kalb_entanglement_2017, humphreys_deterministic_2018,rozpedek_near-term_2019}.

In contrast, the high cooperativity regime $C > 1$ corresponds to conditions of near-deterministic interactions, where the emitter has a high probability to interact with a cavity photon before it dephases. 
High cooperativity is a prerequisite for 
protocols involving deterministic, cavity-mediated quantum information processing with spins and photons \cite{cirac_quantum_1997, duan_scalable_2004, borregaard_quantum_2019}. 
For example, in the high-cooperativity regime, 
single, indistinguishable cavity photons can be generated on demand \cite{mckeever_deterministic_2004}. Since this interaction is coherent and reversible, single photons injected into the cavity can also be completely absorbed by the emitter, enabling transduction of quantum states from light to matter \cite{cirac_quantum_1997, tanji-suzuki_chapter_2011, reiserer_cavity-based_2015}. Alternatively, the spin state of a single quantum emitter can fully modulate the amplitude or phase of a photon reflected from the cavity, 
enabling deterministic interactions with (or among) transient photons \cite{duan_scalable_2004, waks_dipole_2006, waks_dispersive_2006}. 
Finally, strongly-coupled cavity photons can be used to mediate spin-spin interactions in a cavity or between distant, resonant cavities, enabling implementation of near-deterministic distributed quantum logic operations \cite{pellizzari_decoherence_1995, cirac_quantum_1997, imamoglu_quantum_1999, jiang_quantum_2009, monroe_large-scale_2014}. 
Notably, the efficiency and fidelity of such near-deterministic protocols can generally be improved by increasing $C$, motivating the development of cQED systems that can reach $C \gg 1$ \cite{borregaard_quantum_2019}.

A final consideration in cavity engineering of solid-state emitter properties is the efficient in- and out-coupling of light. In particular, the cavity-confined mode must be engineered to 
%
critically- or over-couple into a propagating mode that can be guided into a single-mode fiber with high efficiency [Fig.~\ref{fig:levels}(a)]. 
This is equivalent to the condition $\kappa_c \geq \kappa_s$, where $\kappa_c + \kappa_s = \kappa$ and $\kappa_c$ and $\kappa_s$ are the cavity leakage rates into the collected and scattered modes respectively. 
Beyond ensuring efficient photon collection, $\kappa_c \geq \kappa_s$ is a prerequisite for certain deterministic quantum logic operations between spins and photons \cite{duan_scalable_2004, borregaard_quantum_2019}.
Once coupled out of the cavity via $\kappa_c$, the photons should be mode-matched with high overlap $\eta$ into a single-mode fiber, either for detection or for distribution to distant quantum network nodes.

In summary, in order to achieve high cooperativity, the coherent interaction between the emitter and cavity photons must be made stronger than all decoherence mechanisms by maximizing the ratio $Q/V$ while minimizing undesirable dephasing $\gamma$. Accomplishing this is a central challenge of experimental cQED, and requires careful consideration of both cavity and emitter properties.

\section{Choosing an Emitter}
\label{defect_section}

\begin{table*}[tbp]
\begin{center}
 \begin{tabular}{|c c c c c c|} 
 \hline
 Defect & Symmetry & ZPL wavelength & DW Factor ($\xi$) & lifetime ($\tau$) & 
 $\hbar\Delta_{GS}/k_b\ ^\dagger$  \\ [0.5ex] 
 \hline\hline
 NV & C$_{3v}$& 637 nm & 0.03 \cite{barclay_hybrid_2011} & 11-13 ns \cite{robledo_control_2010,doherty_nitrogen-vacancy_2013}& N/A \\ 
 \hline 
 SiV & D$_{3d}$& 737 nm & 0.7 \cite{dietrich_isotopically_2014} & 1.6-1.7 ns \cite{jahnke_electronphonon_2015,evans_narrow-linewidth_2016,rogers_multiple_2014} & $2.4$ K  \cite{hepp_electronic_2014}\\
 \hline
 GeV & D$_{3d}$&602 nm & 0.6 \cite{palyanov_germanium_2015} & 6 ns \cite{bhaskar_quantum_2017} & $7.3$ K \cite{bhaskar_quantum_2017}\\
 \hline 
 SnV &D$_{3d}$& 619 nm & 0.6 (5K) \cite{gorlitz_spectroscopic_2020} & 4.5-4.8 ns \cite{trusheim_transform-limited_2020, rugar_characterization_2019} & $41$ K \cite{iwasaki_tin-vacancy_2017} \\
 \hline
 PbV &D$_{3d}^*$ &520-552 nm \cite{trusheim_lead-related_2019,tchernij_single-photon_2018} &unknown & >3 ns \cite{tchernij_single-photon_2018,trusheim_lead-related_2019} & $200-270$ K \cite{tchernij_single-photon_2018,trusheim_lead-related_2019}\\ [1ex] 
 \hline
 SiV0&D$_{3d}$& 946 nm & 0.9 \cite{rose_observation_2018} & 1.8 ns \cite{rose_observation_2018} & N/A\\ [1ex] 
 \hline
\end{tabular}
   \captionsetup{justification=centering}
\caption{Summary of emitter properties (see text for details). $*$ PbV symmetry is unconfirmed experimentally, $\dagger$ temperature corresponding to exponential suppression of phonon-induced spin dephasing.\label{opt_table}}
\end{center}
\end{table*}

There exist a multitude of crystallographic defects with optical transitions within the bandgap of diamond \cite{zaitsev_optical_2001}, some of which have optically accessible spin degrees of freedom.
An important class of defects comprises an impurity atom and single vacancy \cite{goss_vacancy-impurity_2005}, of which the most studied are the negatively charged nitrogen-vacancy (NV) [Fig. \ref{defect_fig}(a)] and negatively charged silicon-vacancy (SiV) centers [Fig. \ref{defect_fig}(d)]. In addition, several emerging color centers have recently gained traction in the field, such as the neutral charge state of the silicon vacancy (SiV0), as well as negatively charged group-IV defects based on heavier impurities such as the germanium-vacancy (GeV), tin-vacancy (SnV), and lead-vacancy (PbV) centers. An ideal emitter for quantum information applications would combine deterministic fabrication with coherent, bright optical transitions that couple to long-lived spin states. This section provides a survey of such properties as well as other experimental considerations that are relevant in choosing a color center.

\subsection{Fabrication}

Efficient cavity coupling requires accurate emitter placement within the cavity mode as well as high optical coherence. Simultaneously achieving these requirements poses a significant challenge, and has spurred development of advanced techniques for emitter creation~\cite{smith_colour_2019}. 

The best placement accuracy is obtained using ion implantation and annealing \cite{rabeau_implantation_2006,pezzagna_creation_2010,orwa_engineering_2011}. Standard blanket implantation forms a two-dimensional layer of impurities at a depth determined by the acceleration energy \cite{ziegler_stopping_1985}, while three-dimensional precision can be achieved using a focused ion beam (FIB) \cite{meijer_generation_2005,lesik_maskless_2013,tamura_array_2014,schroder_scalable_2017}, or via blanket implantation through a lithographically aligned mask 
\cite{toyli_chip-scale_2010, spinicelli_engineered_2011, bayn_generation_2015,scarabelli_nanoscale_2016}.
Combining shallow masked implantation and diamond overgrowth~\cite{rugar_generation_2020} could further aid 3D localization by limiting implantation straggle. 
Subsequent high-temperature, high-vacuum annealing repairs lattice damage and mobilizes vacancies to form the desired color center with sub-unity conversion efficiency \cite{pezzagna_creation_2010,naydenov_increasing_2010,chu_coherent_2014}. 
Particularly for larger implanted species, annealing at higher temperatures ($>1200^{\circ}$ C) or pressures may be important to mitigate unwanted defects formed due to implantation damage ~\cite{lang_long_2020,iwasaki_tin-vacancy_2017}.

In contrast to implantation, fabrication techniques based on as-grown impurities offer less control over position but generate defects with better optical properties. This was clearly illustrated by a recent study comparing as-grown and implanted NV centers in the same sample, where the former displayed superior optical coherence [Fig. \ref{defect_fig}(b) \cite{van_dam_optical_2019}]. As-grown NV centers can be formed using the non-negligible native nitrogen impurity levels present in electronic grade diamond. Other impurities can be introduced in high pressure, high temperature (HPHT) \cite{palyanov_germanium_2015,palyanov_high-pressure_2016,palyanov_high-pressure_2019} or chemical vapor deposition (CVD) diamond synthesis \cite{rogers_multiple_2014,bray_single_2018}, and delta-doping techniques have been used to further localize emitters into a single layer \cite{ohno_engineering_2012,lee_deterministic_2014}. 
In addition, precise boron doping has been critical in engineering the Fermi level of diamond to stabilize the SiV0 charge state \cite{rose_observation_2018}. 
Following impurity incorporation, techniques such as electron irradiation \cite{martin_generation_1999,campbell_radiation_2000, mclellan_patterned_2016} or laser writing \cite{chen_laser_2017} can be used to generate vacancies, which, upon annealing, can recombine with implanted or as-grown impurities to form emitters. 

\subsection{Optical Properties}

Cavity-coupled quantum information technologies require a high rate of emission on a coherent optical transition. For diamond defects, such transitions lie within the ZPL. Moreover, to eliminate thermal broadening of the ZPL itself, experiments must be conducted at cryogenic temperatures (typically at $\sim 10$ K or below~\cite{fu_observation_2009, jahnke_electronphonon_2015}). 

The ZPL radiative emission rate ($\gamma_0$) is determined by a combination of the excited state state lifetime ($\tau$), Debye-Waller factor ($\xi$), and quantum efficiency ($QE$) according to 
\begin{equation}
    \gamma_0=\frac{\xi }{\tau}QE,
\end{equation} 
neglecting for simplicity any fine structure within the ZPL (see Table \ref{opt_table} for a comparison of emitter properties). While emitter lifetimes vary by less than an order of magnitude, the Debye-Waller factor, or the fraction of radiative emission that occurs within the ZPL, is much lower for the NV than for group-IV emitters \cite{gali_ab_2013,sipahigil_quantum_2017}. This is due to a change in the NV electronic wavefunctions (or charge distributions) between the ground and excited states, such that photon emission is accompanied by a significant shift in nuclear spin coordinates.

Quantum efficiency refers to the radiative fraction of total excited state decay, which can also include direct phonon relaxation
~\cite{toyli_measurement_2012,jahnke_electronphonon_2015}. In general, $QE$ is challenging to extract directly from emitter brightness due to confounding factors that can reduce fluorescence detection rates, including the presence of metastable dark states and imperfect calibration of collection and detection efficiency. Instead, $QE$ can be most precisely estimated by measuring the response of the emitter's excited state lifetime to a controlled change in the local photonic density of states \cite{buchler_measuring_2005}.
This technique has been used to show that the NV $QE$ is close to unity in bulk diamond \cite{radko_determining_2016}. 
Unfortunately, this method is not as precise for emitters with $QE$ substantially less than $1$, since their lifetimes do not depend as sensitively on their local photonic environment.
This is the case for the SiV center \cite{evans_narrow-linewidth_2016}, which is believed to have $QE\approx 0.1$ at 4 K (see supplementary of \cite{sipahigil_integrated_2016}). Indeed, the SiV's relatively short and temperature-dependent 
lifetime of $\sim 1.6 (1.0)$ ns at $4 (300)$ K \cite{jahnke_electronphonon_2015} is consistent with strong nonradiative processes. On the other hand, the GeV center has a slightly longer, temperature-independent 
lifetime of $6$ ns that is 
very sensitive to its local photonic environment \cite{bhaskar_quantum_2017}, and single GeVs can induce 
coherent extinction of waveguide transmission \cite{bhaskar_quantum_2017, wan_large-scale_2020}. These measurements suggest a relatively high $QE \gtrsim 0.4$. However, this estimate is in conflict with a $QE < 0.1$ extrapolated using detected count rates from 
GeV centers in bulk diamond \cite{chen_optical_2019}. 
There is even more uncertainty regarding the $QE$ of other emerging color centers including the SnV, PbV, and SiV0; 
however, it is worth noting that photon count rates in experiments involving SnV and SiV0 are consistent with a high $QE$ comparable to that of the NV center \cite{iwasaki_tin-vacancy_2017, rose_observation_2018}.

\begin{figure*}[ht!]
\begin{center}
	\includegraphics{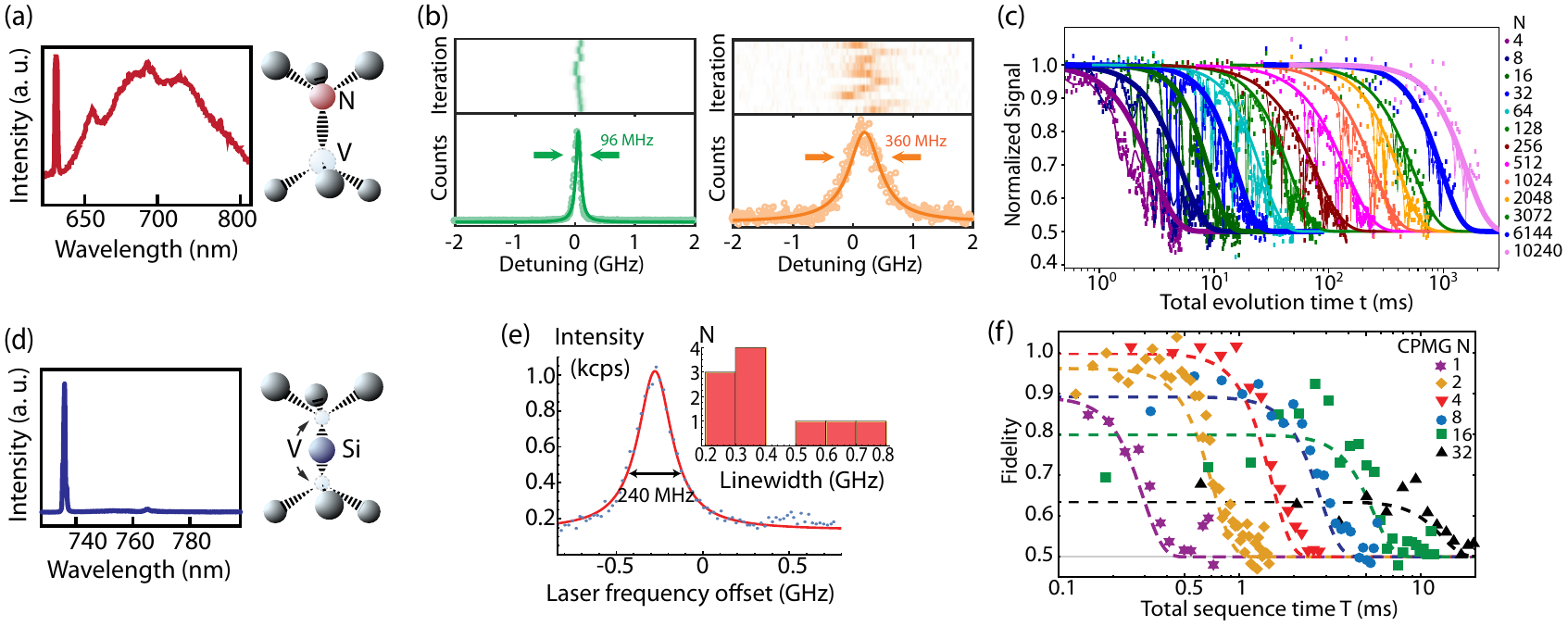}
	\caption{(a) NV center structure and low temperature emission spectrum (spectrum adapted with permission from \cite{jelezko_single_2006}).
	(b) Representative NV photoluminescence excitation (PLE) data at 4 K for each nitrogen isotope (green: $^{\text{14}}$NV; orange: $^{\text{15}}$NV). The sample is implanted with $^{\text{15}}$N, and the $^{\text{14}}$N are as-grown impurities. Individual scans of the ZPL reveal the linewidth free from spectral diffusion. The summation of many repeated scans shows spectral diffusion (adapted with permission from \cite{van_dam_optical_2019}). 
	(c) Decoherence of an NV electronic spin at 3.7 K with tailored decoupling sequences employing pulse numbers from N = 4 to N = 10,240 (adapted with permission from \cite{abobeih_one-second_2018}).  
	(d) SiV center structure and low temperature emission spectrum. (e) Linewidth of a representative implanted SiV at 4 K inside a nano-waveguide measured by PLE spectroscopy (blue points: data; red line: Lorentzian fit). Inset: histogram of emitter linewidths in nanostructures. Most emitters have linewidths within a factor of four of the lifetime limit (94 MHz) (adapted with permission from \cite{evans_narrow-linewidth_2016}). 
	(f) Spin coherence of an SiV electronic spin at 100 mK using CPMG sequences with N = $1, 2, 4, 8, 16$, and $32$ pulses. The longest measured T$_2$ time is 13 ms for N = 32 (adapted with permission from \cite{sukachev_silicon-vacancy_2017}). 
	 \label{defect_fig}}
	\end{center}
\end{figure*}

Another consideration in choosing an emitter is the ZPL emission frequency. Working at longer wavelengths simplifies nanofabrication by allowing larger feature sizes with lower sensitivity to surface roughness.
Furthermore, the GeV, SnV, and PbV ZPL wavelengths occur in the 520-620 nm range, where it is challenging to obtain stable, high-power lasers.

In addition, resonant cavity coupling relies on spectral stability of the emitter, which is strikingly different for the NV center compared to group-IV defects owing to their different symmetries. While all of these color centers occur along the $\langle 111 \rangle$ family of crystal axes, the nitrogen of the NV center sits in place of a missing carbon atom, resulting in a defect with C$_{3v}$ symmetry. Its lack of inversion symmetry permits inequivalent electric dipole moments in the ground and excited states; consequently, the NV center ZPL frequency is strongly impacted by electric field noise on nearby surfaces, causing spectral diffusion, or variation in frequency over time, particularly when illuminated by the green light used to reinitialize the negative NV charge state~\cite{santori_nanophotonics_2010,robledo_control_2010}. This effect is especially severe for implanted NV centers in nanostructures, which typically exhibit spectral diffusion of many GHz, far beyond the $\sim 15$ MHz lifetime limit \cite{faraon_coupling_2012}. Encouragingly, NV centers formed from native nitrogen impurities and electron irradiation have achieved spectral diffusion linewidths of $<250$ MHz in a few-microns-thick diamond membrane \cite{ruf_optically_2019}. Moreover, by applying pulses of green light until the NV transition matches a desired frequency \cite{robledo_high-fidelity_2011}, the effects of pump-induced spectral diffusion can be mitigated, 
and the majority of single-scan linewidths are below $100$ MHz in such membrane samples \cite{ruf_optically_2019}. Conversely, the electric field sensitivity of the NV can be viewed as a resource for tuning the ZPL frequency via the DC Stark effect, which has been used to actively compensate for both spectral diffusion (effectively reducing $\gamma_d$) and spectral mismatch of different defects \cite{tamarat_stark_2006,bassett_electrical_2011, acosta_dynamic_2012}.

In contrast to the NV center, 
group-IV defects take a split-vacancy configuration described by the point group $D_{3d}$, which includes inversion symmetry, leading to a vanishing permanent electric dipole moment. Such defects are insensitive to surface noise to first order; indeed, both as-grown \cite{rogers_multiple_2014} and implanted \cite{evans_narrow-linewidth_2016} SiV centers can display nearly lifetime-limited linewidths, even in nanostructures \cite{evans_narrow-linewidth_2016,schroder_scalable_2017} [Fig. \ref{defect_fig}(e)]. 
Other group-IV color centers such as the GeV \cite{siyushev_optical_2017,bhaskar_quantum_2017}, SnV~\cite{trusheim_transform-limited_2020}, and SiV0 \cite{rose_observation_2018} exhibit similar spectral stability, although this has not yet been observed in nanofabricated cavities. The insensitivity of group-IV emitters to electric fields precludes Stark shift tuning of the ZPL frequency; instead, two photon Raman transitions \cite{sipahigil_integrated_2016, sun_cavity-enhanced_2018} and dynamic control of the strain environment \cite{meesala_strain_2018, machielse_quantum_2019} are promising approaches for wavelength tuning and spectral stabilization. 

Finally, it is desirable to minimize inhomogeneous broadening, or the variation in ZPL emission frequency from emitter to emitter. The inhomogeneous distribution scales with implantation damage, increasing with the size and energy of the ion, but can be mitigated to varying degrees via post-implantation treatment. 
NV and SiV centers can exhibit inhomogeneous linewidths down to $0.17$ nm \cite{orwa_engineering_2011} and $0.03$ nm \cite{evans_narrow-linewidth_2016} respectively when annealed at high temperatures. 
Much larger linewidths of $30$ nm have been observed for implanted SnV centers \cite{trusheim_transform-limited_2020}, although subsequent high-pressure, high-temperature annealing was shown to achieve distributions down to 6 nm \cite{iwasaki_tin-vacancy_2017}. Furthermore, the spectral features attributed to the PbV ZPL around $520$ nm exhibit a narrow distribution of only 0.12 nm; however, 
unidentified emission lines 
over a $\sim100$ nm range could be evidence of a much larger inhomogeneous distribution or intermediate defect formation 
~\cite{trusheim_lead-related_2019}, meriting further investigation. In practice, it is likely that a combination of improved inhomogeneous broadening and spectral tuning will be necessary to realize spectrally indistinguishable emitters.

\subsection{Spin Properties}

Many quantum information applications rely on spin-photon transduction via state-selective optical transitions within the ZPL \cite{doherty_nitrogen-vacancy_2013,muller_optical_2014,siyushev_optical_2017,rugar_characterization_2019,trusheim_transform-limited_2020}. These spin states, in combination with proximal nuclear spins coupled by magnetic dipolar interactions, can serve as an additional resource for storing or processing information \cite{dutt_quantum_2007} and performing local error correction \cite{taminiau_universal_2014}.

In practice, the NV and SiV0 spins are easiest to work with due to their orbital singlet, $\mathcal{S}=1$ ground states. A
spin qubit can be 
realized between the $m_s=0$ and either of the $m_s=\pm1$ spin states, which are naturally separated in energy by a zero-field splitting. Coherent spin manipulation can be achieved 
via microwave fields~\cite{jelezko_observation_2004}, and NV centers have demonstrated the longest coherence times for a single electron spin qubit in any system (T$_2>1$ s)
[Fig. \ref{defect_fig}(c)] \cite{abobeih_one-second_2018}. 
So far, only ensemble spin resonance has been demonstrated for the novel SiV0 center; nevertheless, these defects exhibit coherence times as long as T$_2=255$ ms at 4 K \cite{rose_observation_2018,rose_strongly_2018,zhang_optically_2020}.

In contrast, color centers based on negatively-charged group-IV defects exhibit a doubly-degenerate ground state in both orbit and spin ($\mathcal{S}=1/2$), with orbital degeneracy lifted by $\Delta_{GS}$ due to a combination of spin-orbit interaction and dynamic Jahn-Teller effect \cite{hepp_electronic_2014,thiering_ab_2018}. 
In an external magnetic field aligned with the $\langle 111 \rangle$ axis, the lowest energy spin-1/2 manifold can be addressed using highly cycling, spin-selective optical transitions~\cite{hepp_electronic_2014, muller_optical_2014,rogers_all-optical_2014, pingault_all-optical_2014}.
%
The major challenge in working with the spins of negatively charged group-IV defects is rapid ground state dephasing caused by single-phonon transitions between orbital states. 
This motivates qubit operation at temperatures well below $\hbar\Delta_{GS}/k_B$ (see Table \ref{opt_table}) to reduce phonon occupation, thereby exponentially increasing spin coherence times. As a result, the SiV exhibits a 4-5 order of magnitude increase in spin coherence times at T $< 500$ mK [T$_2 > 10$ ms \cite{sukachev_silicon-vacancy_2017}, Fig. \ref{defect_fig}(f)] compared to 4 K (T$_2 \sim $100 ns \cite{pingault_coherent_2017}).
A complementary approach for improving spin coherence involves increasing the ground state splitting through the application of strain \cite{sohn_controlling_2018,meesala_strain_2018}. For instance, strain tuning of SiVs in nanostructures has demonstrated an order-of-magnitude increase in orbital splitting, resulting in the highest reported SiV coherence time of T$_2=250$ ns at 4 K \cite{sohn_controlling_2018}. Moreover, defects based on heavier group-IV ions exhibit larger $\Delta_{GS}$, which could facilitate operation at higher temperatures. Indeed, the PbV orbital splitting is estimated to be in the THz regime, suggesting the possibility of long-lived spin coherence at 4 K \cite{trusheim_lead-related_2019,tchernij_single-photon_2018}.

A final consideration is the ability to couple to proximal nuclear spins of either the defect impurity or $^{13}$C carbon isotopes in diamond.
Impressively, the NV center has been used to control a $10$-qubit quantum register with coherence times of > $75$ s \cite{bradley_ten-qubit_2019}. 
Single-site nuclear spin manipulation has also been demonstrated with the SiV \cite{metsch_initialization_2019,nguyen_quantum_2019},
but multi-nuclear-spin registers have not yet been realized.

\subsection{Discussion}
Thus far, state-of-the-art cavity experiments with diamond defects have used either the NV or SiV. While the NV center is the best understood defect, exhibiting excellent spin coherence, its optical properties are poor. In particular, the difference in permanent electric dipole moments between the ground and excited states degrades optical coherence for near-surface emitters. Open Fabry-Perot microcavities containing bulk-like diamond membranes are therefore especially promising for these color centers as they can be situated far from interfaces (see Sec. \ref{fpcav}). In contrast, the SiV has poor spin properties at 4 K, requiring operation at dilution refrigerator temperatures to realize a long-lived spin qubit, but has superior optical coherence. Crucially, its inversion symmetry inhibits sensitivity to surfaces, allowing for incorporation into heavily fabricated nanoscale resonators with high $Q/V$ (see Sec. \ref{nanophotonics}).

In addition, we discussed the potential of emerging group-IV color centers. Negatively-charged group-IV emitters based on larger ions appear to share the attractive optical properties of the SiV with the added potential for improved spin coherence times at 4 K, although these emitters are harder to fabricate and are not as well understood. The recently discovered SiV0 could potentially combine excellent optical and spin properties at liquid helium temperatures, but it requires specially-doped diamond to stabilize the neutral charge state and is one of the least explored of the defects considered here. Finally, recent progress in ab initio \cite{thiering_ab_2018,ciccarino_strong_2020,harris_group_2019} and machine learning \cite{butler_machine_2018, schmidt_recent_2019} techniques suggests it may soon be possible to predict new emitters with superior properties to those discussed in this section.


\section{Open Fabry-Perot Microcavities} \label{fpcav}

Early cQED experiments were performed using cm-scale Fabry-Perot cavities with Gaussian modes defined by two spherical mirrors \cite{haroche_short_2007,kimble_strong_1998}. Such cavities take advantage of absorption-limited dielectric mirror coatings to achieve very high reflectivity \cite{rempe_measurement_1992,hood_characterization_2001}, and offer full spatial and spectral tunability by positioning the mirrors, but their cm-scale size leads to large mode volumes and limited scalability. 
By miniaturizing the parabolic mirrors, these drawbacks can be mitigated, with micron scale geometries achieving mode volumes on the order of $\lambda^3$~\cite{bitarafan_-chip_2017}. 
Moreover, the small size of the micromirrors enables parallelized fabrication processes \cite{dolan_femtoliter_2010,derntl_arrays_2014,bitarafan_-chip_2017} and direct integration with optical fibers \cite{hunger_fiber_2010,muller_ultrahigh-finesse_2010}. 

So far, open microcavities have been coupled to a multitude of quantum systems including atoms \cite{gallego_strong_2018}, ions \cite{steiner_single_2013,kobel_deterministic_2020}, molecules \cite{wang_coherent_2017,wang_turning_2019}, quantum dots \cite{najer_gated_2019}, rare earth ions \cite{casabone_dynamic_2020}, optomechanical systems \cite{flowers-jacobs_fiber-cavity-based_2012}, and color centers in both nanodiamonds \cite{albrecht_coupling_2013,kaupp_scaling_2013,albrecht_narrow-band_2014,johnson_tunable_2015,kaupp_purcell-enhanced_2016,benedikter_cavity-enhanced_2017} and diamond membranes \cite{riedel_deterministic_2017,hausler_diamond_2019,hoy_jensen_cavity-enhanced_2020,salz_cryogenic_2020}. 
Early results with diamond defects utilized nanodiamonds containing single NV centers at elevated temperatures \cite{albrecht_coupling_2013,albrecht_narrow-band_2014, johnson_tunable_2015} to observe funneling of the broad emission into the narrow resonator mode (see Sec. \ref{cqed}).
However, NVs in nanodiamonds are poorly suited to spin-photon interfaces, as their optical properties are degraded by nearby surfaces, motivating the development of a membrane-in-cavity geometry in which emitters can exhibit coherent optical transitions at 4 K \cite{ruf_optically_2019}.
Nevertheless, despite substantial progress in both mirror and diamond fabrication, obtaining $C>1$ for a color centers in a membrane-in-cavity system has not yet been achieved. In this section, we outline the theoretical and experimental progress toward this goal, followed by a near-term outlook for the field. 

\begin{figure}[tbp]
\begin{center}
	\includegraphics{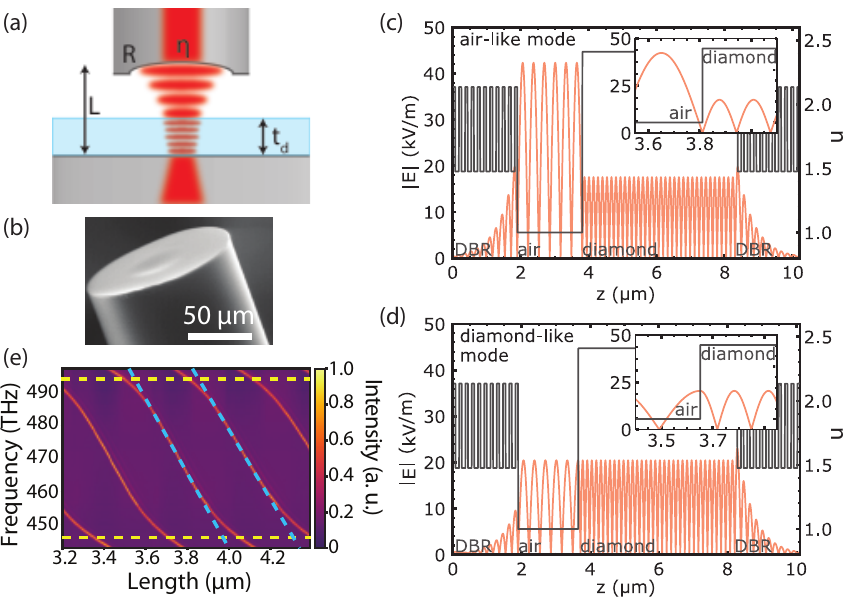}
	\caption{(a) Schematic of a fiber cavity (figure adapted with permission from \cite{janitz_fabry-perot_2015}).
	(b) Scanning electron microscope (SEM) image of a laser-machined fiber tip (adapted with permission from \cite{hunger_fiber_2010}).
	c/d) The refractive index $n$ (black, right axis)
	and electric field strength (orange, left axis) for (c) an air-like mode and (d) a diamond-like mode (adapted from \cite{van_dam_optimal_2018}).
	(e) Membrane-in-cavity mode structure, exhibiting large avoided crossings between diamond (yellow) and air (blue) modes (adapted with permission from \cite{hoy_jensen_cavity-enhanced_2020}).
	\label{open_cav_1}}
	\end{center}
\end{figure}

\subsection{Geometry}
For coupling to defects in a diamond membrane, most experiments employ a half-symmetric geometry [Fig. \ref{open_cav_1}(a)] formed by a microscopic spherical mirror with radius of curvature $R$, fabricated either on the tip of an optical fiber or planar substrate [Fig. \ref{open_cav_1}(b)]. Diamond defects are contained within a membrane of thickness $t_d$ bonded to the macroscopic flat mirror. The spacing between the spherical and flat mirrors determines the cavity length $L$.

As discussed in Sec. \ref{cqed}, it is desirable to obtain a high ratio of cavity quality factor $Q$ to mode volume $V$. 
The mode volume is $V\approx \pi \omega_0^2 L_{\text{eff}}/4$, where $\omega_0$ is the cavity waist and $L_{\text{eff}}$ is the effective cavity length expressed in terms of an equivalent distance in diamond \cite{van_dam_optimal_2018}, $L_\text{eff} = \left(\nicefrac{2}{n_d^2 |E_{d}|^2}\right)\int_{cav}n^2(z)|E(z)|^2 dz$. 
Here, $n(z)$ and $E(z)$ are the refractive index and electric field at position $z$ in the cavity, and $n_d$ and $E_{d}$ are the index and maximum electric field in diamond. However, in principle, $Q/V$ can be largely independent of $L_\text{eff}$ because the photon lifetime -- and thus $Q$ -- typically increases linearly with cavity length when losses occur at mirrors and interfaces. Consequently, the figure of merit becomes $Q/V
\propto \mathcal{F}/(\lambda\omega_0^2)$, where 
the finesse $\mathcal{F}$ is $\pi$ divided by the effective per-pass losses in the cavity.
Nevertheless, in practice it is easiest to simultaneously achieve small waist $\omega_0$ and high finesse $\mathcal{F}$ using short microcavities (see discussion of losses below). Consequently, maximizing $Q/V$ for open-geometry cavities typically involves minimizing losses and reducing $R$ (since $\omega_0\propto\sqrt{R}$ for $R\gg L$), as well as $L$, $t_d$, and electric field penetration into the mirror coatings.

A final geometric consideration is collection efficiency, which is determined by the overlap $\eta$ between the transmitted cavity electric field and that of a single mode fiber. 
Micromirrors fabricated on planar substrates allow introduction of optical elements to maximize fiber coupling of outcoupled light \cite{greuter_small_2014}, but in turn require challenging optical routing in a cryostat, although novel alignment techniques may alleviate some of this difficulty \cite{riedel_cavity-enhanced_2020}.
In contrast, mirrors fabricated directly on the tips of optical fibers feature direct coupling from the cavity to the propagating fiber mode. This coupling is maximized by precisely centering the mirror relative to the fiber core, while matching the fiber and cavity mode diameters and minimizing wavefront curvature \cite{hunger_fiber_2010,gallego_high-finesse_2016}. 
For such cavities, there exists a trade-off between obtaining a small mode waist (requiring small $R$) and collection efficiency through the fiber mirror due to wavefront curvature mismatch.
Nevertheless, power coupling efficiencies of over $85\%$ have been demonstrated with empty fiber cavities \cite{hunger_fiber_2010}. 

\subsection{Mode Structure}
\label{open_cav_mode}
The diamond membrane strongly modifies the cavity mode structure and losses compared to those of an empty or "bare" cavity. This behavior is well captured by a
simplified one-dimensional model of the cavity electric field 
with nodes at the mirror interfaces \cite{janitz_fabry-perot_2015,van_dam_optimal_2018}.
The field at the air-diamond interface is therefore completely determined by the membrane thickness $t_d$, which sets the relative energy density in the air and diamond regions via the electromagnetic boundary conditions.
A field node at this interface leads to an "air-like" mode, where the intensity is higher in the air versus the diamond region by a factor of $n_d$ [Fig. \ref{open_cav_1}(c)], while an anti-node leads to a "diamond-like" mode where the opposite is true [Fig. \ref{open_cav_1}(d)]. 

The mode types are readily identified from cavity transmission spectra 
[Fig. \ref{open_cav_1}(e)]. The canted periodic structure can be understood by considering the limit of a perfectly reflective air-diamond interface (i.e. $n_d \rightarrow \infty$), in which case the cavity is divided into "air" and "diamond" modes [blue and yellow lines in Fig. \ref{open_cav_1}(e)]. The frequency spacing of these modes depends on the diamond thickness and cavity length according to $\Delta \nu_{di} = c/(2 n_d t_d)$ and $\Delta \nu_{air} = c/(2 ( L - t_d))$. With finite $n_d$ these modes are coupled to one another, leading to the large avoided crossings observed in the spectrum, where diamond-like modes have a shallow slope, and air-like modes have a steeper slope \cite{janitz_fabry-perot_2015}. The sensitivity of field localization to frequency and diamond thickness offers an extra layer of tunability to the open-cavity geometry. 

Based on available mirrors, open cavities can in principle achieve an extraordinary finesse of $\sim 10^5$ or more \cite{rempe_measurement_1992, hood_characterization_2001}.
In practice, it is very challenging to achieve such high values in membrane-in-cavity systems. Losses due to scattering or absorption at the air-diamond interface display exquisite sensitivity to the field amplitude at this position; such losses are consequently maximized for diamond-like modes and minimized for air-like modes \cite{janitz_fabry-perot_2015}. Models including known surface roughness \cite{janitz_fabry-perot_2015,van_dam_optimal_2018} exhibit smaller surface losses than seen in experiments \cite{janitz_fabry-perot_2015,riedel_deterministic_2017}, indicating significant contributions from absorption at the diamond surface. In contrast, bulk membrane absorption can be neglected for electronic grade samples \cite{friel_development_2010}. 

\begin{figure}[!b]
\begin{center}
\includegraphics{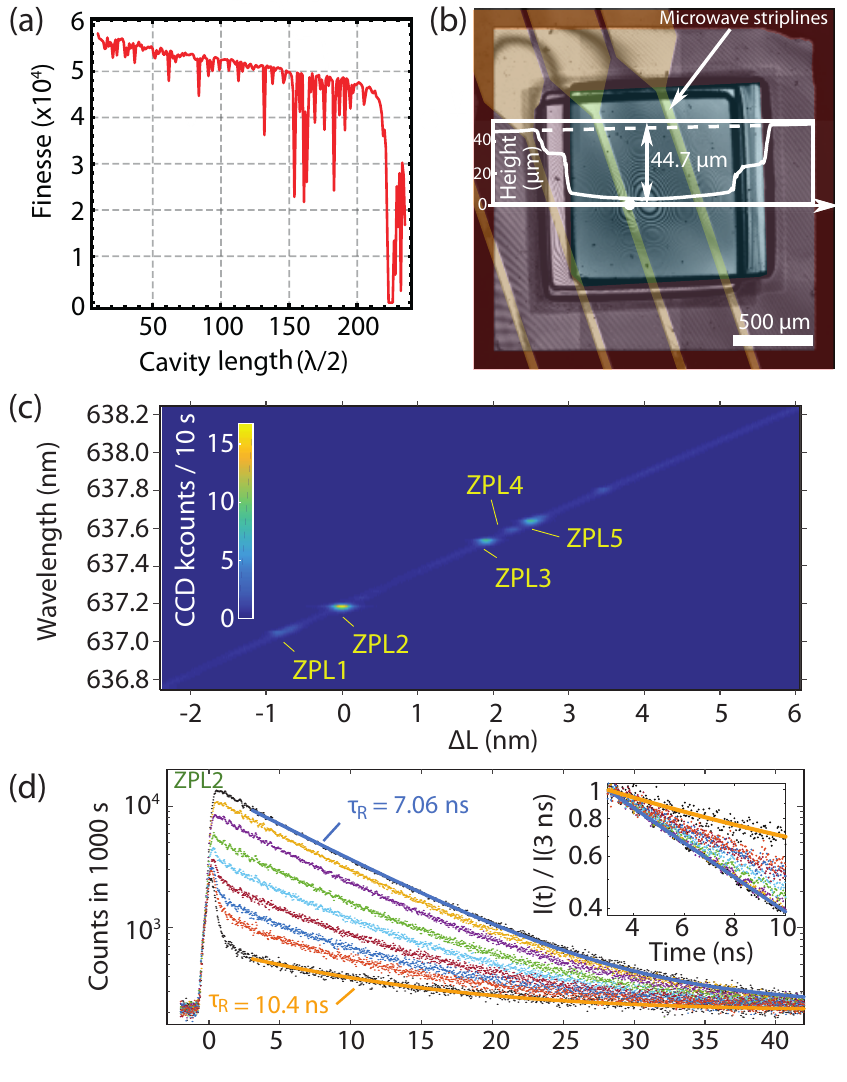}
\caption{(a) Cavity finesse as a function of axial mode order. Drops in finesse arise from perturbative coupling to lossy higher order modes caused by the Gaussian shape of fiber mirror (adapted with permission from \cite{benedikter_transverse-mode_2015}). 
(b) Confocal microscope image of an etched diamond membrane with false color added. The white arrow indicates the path along which the height profile was measured (adapted with permission from \cite{ruf_optically_2019}).
(c) PL spectra around the NV ZPL transition frequency for different air-gap detunings $\Delta L$. 
Each resonance corresponds to the ZPL emission of a single defect. (d) Photoluminescence decay curves of "ZPL2" from (c) following pulsed excitation at different cavity lengths.
The inset shows the normalized decay curves corresponding to a reduction in lifetime from $12.6$ to $7.06$ ns (Figs. (c) and (d) were adapted with permission from \cite{riedel_deterministic_2017}).
\label{fig_cav2} }
\end{center}
\end{figure}

Other sources of loss stem from non-ideal cavity geometry, as the finite extent of the micromirror can lead to clipping losses for large mode diameters (which increase with cavity length) \cite{hunger_fiber_2010}. Such losses are exacerbated when the cavity mode couples to higher-order transverse modes, which can result from deviation of the micromirror from an ideal parabola \cite{benedikter_transverse-mode_2019} [Fig. \ref{fig_cav2}(a)] and deviation of the flat mirror from an ideal plane \cite{benedikter_transverse-mode_2015}.

\subsection{Fabrication}
Scalable quantum technologies based on open microcavities require deterministic and repeatable fabrication processes for both the micromirror and diamond membrane. Micromirror fabrication seeks to produce shallow parabolic dimples with micron-scale radius of curvature, while maintaining near atomically smooth surfaces compatible with high finesse dielectric coatings. So far, the most popular methods for machining mirror templates are based on laser ablation, FIB, and silicon etching. Laser ablation using a CO$_2$ laser results in controlled creation of Gaussian-shaped dimples with low surface roughness ($0.2$ nm-rms) \cite{nowak_efficient_2006,hunger_laser_2012,ruelle_optimized_2019}. Recently, the inclusion of additional nanofabrication steps achieved an effective $R<5\ \upmu$m (based on the frequency spacing between higher-order transverse modes),
with ablation depths of only $\approx1\ \upmu$m \cite{najer_fabrication_2017}. In contrast, FIB milling has been used to fabricate dimples with very small radius of curvature and precisely controlled geometry at the expense of higher surface roughness ($0.3-0.8$ nm-rms \cite{dolan_femtoliter_2010,albrecht_narrow-band_2014,trichet_topographic_2015}). This method produced dimples with 
effective $R=4.3\ \upmu$m and depths of only $230$ nm \cite{trichet_topographic_2015}. Finally, silicon etching has been used to generate large arrays of ultra-smooth (0.2 nm-rms) mirror templates \cite{biedermann_ultrasmooth_2010,wachter_silicon_2019}. So far, this technique has produced relatively large radii of curvature ($R>100\ \upmu$m), but smaller geometries may be possible. 
 Once fabricated, mirror substrates are subsequently coated with low-loss dielectric mirrors. After deposition, it is possible to integrate microwave striplines for spin control directly into the flat mirror substrate while maintaining a high cavity finesse \cite{bogdanovic_robust_2017} [see Fig. \ref{fig_cav2}(b)].

The second fabrication requirement is a process capable of producing ultra-smooth, micron-thick diamond membranes containing individual defects
with bulk-like optical properties. The starting point for such a membrane is typically a commercially available $\sim100\ \upmu$m-thick electronic grade diamond plate. Etching hundreds of microns to obtain a single membrane constitutes a long and wasteful process; consequently, membrane substrates are generally obtained from bulk samples either by laser slicing and polishing (resulting in $t_d\sim10\ \upmu$m) or via a novel implantation process 
~\cite{piracha_scalable_2016,salz_cryogenic_2020}. The latter technique involves implanting He atoms to form an amorphous subsurface layer, which is converted to graphite by subsequent annealing. The diamond is then overgrown with pristine single crystal diamond, and released at the graphitic layer with electrochemical etching, yielding $t_d\sim100$ nm -$1\ \upmu$m. Both techniques produce a surface roughness of a few nm, but thus far membranes based on laser-sliced substrates have achieved the best results for surface losses \cite{hoy_jensen_cavity-enhanced_2020} and fabrication of individually-addressable defect centers \cite{riedel_deterministic_2017,ruf_optically_2019,hoy_jensen_cavity-enhanced_2020} (see Sec. \ref{defect_section}). Membrane substrates can then be bonded to carrier wafers via van der Waals forces for further processing. Thinning and smoothing of both surfaces is achieved through an inductively coupled plasma reactive ion etching (ICP RIE) process, which cycles between ArCl$_2$ and O$_2$-based recipes \cite{ovartchaiyapong_high_2012,tao_single-crystal_2014,latawiec_-chip_2015,appel_fabrication_2016}, obtaining a final surface roughness as low as $\sim0.1$ nm-rms \cite{sangtawesin_origins_2019}. Once the desired device thickness is achieved, the sample can be transferred using a micromanipulator \cite{riedel_deterministic_2017} or in a water droplet to the flat mirror substrate. Recent results show that it is also possible to conduct the final etch through a quartz mask after bonding the sample to the mirror, reducing membrane handling \cite{ruf_optically_2019} [Fig. \ref{fig_cav2}(b)].

\subsection{State of the art}
Despite considerable experimental progress, there has been only one demonstration of a single emitter in a membrane coupled to an open microcavity at low temperature \cite{riedel_deterministic_2017}. In this experiment, an implanted NV center was coupled to a $\mathcal{F}=5,260$ air-like mode [Fig. \ref{fig_cav2}(c)] resulting in a ZPL enhancement of $\mathcal{P}\approx30$, an excited-state lifetime reduction by a factor of $2.0$, and $46\%$ of photons emitted into the ZPL. 
Furthermore, the tunability of the system was illustrated by a variation in excited-state lifetime as a function of cavity length [Fig. \ref{fig_cav2}(d)]. While this experiment represents a major milestone in coupling open cavities to diamond defects, it achieved an estimated cooperativity of only $C\approx0.03$. This value is strongly impacted by an increase in spectral diffusion from 100 MHz to 1 GHz associated with thinning the membrane to $t_d<1\ \upmu$m, and could potentially be improved by an order of magnitude by using a slightly thicker sample containing NVs formed by native nitrogen and electron irradiation \cite{van_dam_optical_2019,ruf_optically_2019} or by using more stable group-IV defects. Furthermore, reduced losses at the diamond interface would increase the cavity finesse and permit
 coupling to diamond-like modes. Such improvements recently led to the coupling of a single GeV center to a finesse $11,000$ diamond-like mode of an open microcavity \cite{hoy_jensen_cavity-enhanced_2020}.
Finally, cavity mode volume could be reduced by decreasing the mirror radius of curvature using a combination of laser ablation and nanofabrication techniques \cite{najer_fabrication_2017}. 

Beyond fabrication, there are additional technical barriers limiting emitter-cavity coupling strength. Resonant coupling requires stabilization of the cavity length to well within a linewidth, on the order of $10$ pm for state-of-the-art mirrors. For instance, the liquid helium cryostat used in the aforementioned NV experiment induced vibrations of 24 pm-rms in the passively stabilized cavity, representing a significant fraction of the 60 pm cavity linewidth \cite{riedel_engineering_2017}. Consequently, many avenues for increasing stability have been explored such as rigid mounting \cite{gallego_high-finesse_2016}, external vibration isolation \cite{bogdanovic_design_2017}, thermal feedback of the mirror coatings \cite{brachmann_photothermal_2016,gallego_high-finesse_2016}, as well as active locking methods such as Pound-Drever-Hall \cite{gallego_high-finesse_2016,janitz_high_2017}, side-of-fringe \cite{salz_cryogenic_2020}, and H{\"a}nsch-Couillaud techniques \cite{hansch_laser_1980,wang_turning_2019}.

\subsection{Outlook}
Recent experimental milestones suggest that it should soon be possible to couple bulk-like emitters in diamond membranes to high-finesse open microcavities. This is particularly promising for achieving $\mathcal{C}>1$ for an NV center, which has been a long-standing challenge due to the degradation of emitter optical properties in nanofabricated resonators (see Sec. \ref{nanophotonics}). 
Furthermore, the narrow cavity linewidths afforded by open cavities offer an opportunity for spin-selective enhancement, opening the door for high fidelity spin measurement \cite{hanks_high-fidelity_2017}, as well as schemes for quantum communication, computation, and metrology \cite{nemoto_photonic_2014,nemoto_photonic_2016,hanks_universal_2017}. 
Finally, while scaling to multi-cavity systems remains an active area of research~\cite{trupke_large-scale_2016}, progress in parallelized open-cavity fabrication techniques~\cite{derntl_arrays_2014} shows the potential for larger-scale technologies.

\section{Nanophotonic cavities} \label{nanophotonics}

\begin{figure*}[h!]
	\begin{center}
		\includegraphics[scale=1]{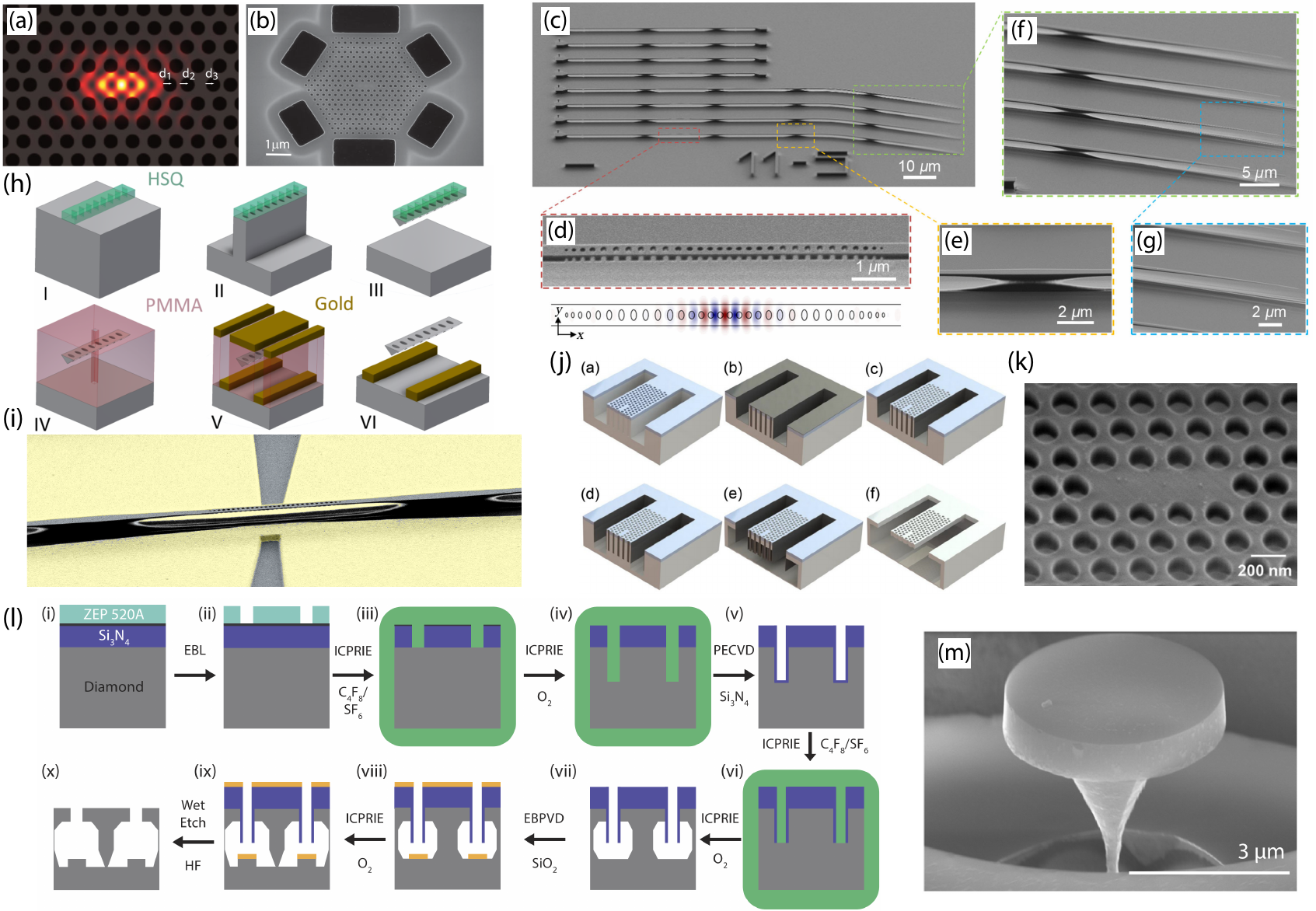}
		\caption{ \label{fig:nano}
			Diamond nanophotonic devices.
			(a) Mode simulation and (b) SEM of 2D PCC fabricated from an ultra-thin diamond membrane \cite{faraon_coupling_2012}. (c) SEM of an array of free-standing PCCs created by angled reactive ion etching, highlighting (d) 1D PCC region, (e) support anchor connecting waveguide to substrate, and (f-g) tapered waveguide region for adiabatic single-mode fiber coupling \cite{burek_fiber-coupled_2017}. (h) Fabrication procedure for angle-etched devices (steps I-III), including targeted implantation through a mask (IV) and microwave electrode deposition (V-VI) \cite{nguyen_integrated_2019}. (i) SEM of resulting devices with gold coplanar waveguide for microwave spin control (false color added, adapted with permission from \cite{nguyen_quantum_2019}). (j) Rectangular cross-section PCC fabrication procedure using an anisotropic crystallographic etch and (k) SEM of a freestanding 2D PCC \cite{wan_two-dimensional_2018}. (l) Optimized procedure to fabricate microdisk resonators using the crystallographic etch and (m) resulting high $Q/V_0$ device \cite{mitchell_realizing_2019}.
		}
	\end{center}
\end{figure*}

In contrast to open-geometry cavities, nanophotonic resonators can readily achieve sub-micron mode volumes by using refractive-index contrast to confine light to volumes of order $(\lambda/n)^3$ or smaller \cite{joannopoulos_photonic_2011, robinson_ultrasmall_2005, hu_design_2016, choi_self-similar_2017}. 
Additionally, nanophotonic structures are naturally desirable for long-term scalability, since they can be fabricated en masse and utilize on-chip photon routing \cite{qiang_large-scale_2018, papon_nanomechanical_2019}.
For a general, platform-agnostic overview of nanophotonic systems we direct the reader to an alternate reference \cite{koenderink_nanophotonics_2015}.

With its high index of refraction $n = 2.4$, diamond is a natural candidate for such systems, and in the absence of scattering or absorption losses, diamond nanophotonic resonators could theoretically achieve $\mathcal P \sim 10^5$ \cite{burek_high_2014}, or $Q/V_0 > 10^6$, where $V_0 = V (n/\lambda)^3$ is the mode volume relative to a cubic wavelength in the diamond.
A wide variety of cQED structures have been fabricated in diamond \cite{aharonovich_diamond_2014}, including whispering gallery mode resonators \cite{khanaliloo_high-qv_2015, mitchell_realizing_2019}, ring resonators \cite{faraon_resonant_2011, hausmann_integrated_2012}, and photonic crystal cavities (PCCs) \cite{wang_fabrication_2007, babinec_design_2011, bayn_processing_2011, riedrich-moller_one-_2012}.
Here, we focus on the unique technical considerations involved in realizing and utilizing nanophotonic structures for cQED in single-crystal diamond, and related experimental progress therein.

\subsection{Fabrication techniques} \label{nanofab}

The typical approach for nanoscale fabrication of high-quality photonic devices begins with single crystal thin films ($\sim 100$ nm) grown heteroepitaxially, such as silicon-on-insulator, which can then be processed into undercut photonic structures using standard lithography and wet-etching techniques \cite{jalali_silicon_2006}.
    Unfortunately, this relatively straightforward approach to engineer optically isolated photonic structures does not translate effectively to the fabrication of diamond. Despite immense progress in the field of diamond growth, heteroepitaxial single-crystal thin films of diamond cannot yet be produced with defects at or below the parts-per-billion level, as required for quantum optical experiments involving single emitters \cite{butler_thin_1993, jiang_heteroepitaxial_1993, schreck_diamondirsrtio3_1999, teraji_chemical_2006, rath_waferscale_2013}. Additionally, diamond is resilient to all forms of wet etching. Instead, less controllable plasma-based dry etching techniques \cite{sandhu_reactive_1989} must be used in combination with sophisticated lithography masks \cite{ burek_free-standing_2016, mitchell_realizing_2019}. Initial successes in nanofabrication of diamond [for example, see Fig.~\ref{fig:nano}(a-b)] overcame these challenges using creative techniques to engineer nanoscale diamond membranes with sub-micron thickness, either via ion-beam implantation and liftoff or mechanical polishing and subsequent reactive-ion etch thinning, but both techniques struggled with low resulting cavity and emitter quality \cite{parikh_singlecrystal_1992, olivero_ion-beam-assisted_2005, wang_fabrication_2007, fairchild_fabrication_2008, faraon_coupling_2012, hausmann_integrated_2012}.

Rather than membrane-based fabrication, underetching of structures defined in bulk, electronic-grade, single-crystal substrates \cite{bayn_triangular_2011} has recently enabled high-quality nanophotonic devices [Fig.~\ref{fig:nano}(c-m)]. Unlike most nanoscale membrane structures, devices undercut from bulk substrates can achieve high quality factors due to their optical isolation from the bulk and are compatible with annealing and acid-cleaning post-processing steps required for incorporation of high-quality single emitters (Sec.~\ref{defect_section}). Underetching was first implemented by using an angled etch to undercut one-dimensional structures predefined with electron-beam lithography and top-down etching, leaving behind freestanding diamond nanobeams with a triangular cross-section [Fig.~\ref{fig:nano}(h), steps I-III]. The angled etch was initially accomplished by placing samples inside a triangular Faraday cage within the reactive-ion etcher \cite{burek_free-standing_2012, burek_high_2014}. More recently, a similar angled etch has been achieved using ion-beam milling at a well-controlled angle of incidence, resulting in more reliable etch profiles \cite{atikian_freestanding_2017}. Both of these techniques have proven extremely effective for fabricating freestanding one-dimensional PCCs out of electronic-grade diamond substrates, such as those shown in Fig.~\ref{fig:nano}(c-i), with high ratios of $Q/V_0 > 10^4$ \cite{burek_high_2014, nguyen_integrated_2019}.

Underetched devices can also be fabricated out of bulk single-crystal diamond using a selective crystallographic etch to achieve a flat lower surface, as illustrated in Fig.~\ref{fig:nano}(l) \cite{khanaliloo_single-crystal_2015, mitchell_realizing_2019}. This etching technique was initially developed to fabricate high $Q/V_0$ whispering gallery mode resonators [Fig.~\ref{fig:nano}(m)] \cite{khanaliloo_high-qv_2015, mitchell_realizing_2019}, 
and has now been adapted to fabricate freestanding PCCs with a rectangular cross-section \cite{mouradian_rectangular_2017}, enabling two-dimensional PCCs in single-crystal diamond [Fig.~\ref{fig:nano}(j-k)] \cite{wan_two-dimensional_2018}. While the crystallographic-etch technique is somewhat less mature than the angled etch, it allows for rectangular device geometries similar to conventional photonic platforms. This should enable fabrication of waveguides with arbitrary relative spacing, allowing for implementation of freestanding diamond-waveguide based beamsplitters and electromechanical switches for on-chip photon routing \cite{qiang_large-scale_2018, papon_nanomechanical_2019}. Furthermore, this technique has a largely unexplored parameter space available, making it possible to optimize the mask and etching procedure to achieve a wide variety of surface properties \cite{kiss_high-quality_2019, mitchell_realizing_2019}.

In order to utilize low-mode-volume cavities for cQED experiments, color centers must be placed at the mode-field maximum of the cavity with sub-$100$ nm precision. This has been accomplished with SiV centers using a variety of techniques (Sec.~\ref{defect_section}), including delta doping \cite{zhang_strongly_2018}, FIB implantation~\cite{tamura_array_2014, schroder_scalable_2017, lesik_maskless_2013} and blanket ion implantation through lithographically aligned masks ~\cite{toyli_chip-scale_2010, spinicelli_engineered_2011, bayn_generation_2015}. By implanting several ions and utilizing spectral selection of individual emitters, these techniques have enabled deterministic nanoscale PCC coupling of single SiV centers \cite{sipahigil_integrated_2016, evans_photon-mediated_2018, nguyen_quantum_2019}. 

Once fabricated, nanoscale cavities can be optically interrogated using free-space optics \cite{faraon_resonant_2011, faraon_coupling_2012} or a transient coupler such as a tapered fiber \cite{hausmann_integrated_2013, burek_high_2014}. Coupling can also be made indirectly by means of a waveguide in which the photonic crystal cavity is integrated [for example, see Fig.~\ref{fig:nano}(c, d, i)]. In this case, the cavity design is adjusted to damp preferentially into the diamond waveguide mode (see Sec.~\ref{cqed}, \cite{joannopoulos_photonic_2011}), which can subsequently be outcoupled into a single-mode fiber by various techniques. Even a small defect in the waveguide, such as a deliberately introduced notch, induces scattering into free-space modes that can be coupled into a single-mode fiber via a high-numerical-aperture objective, albeit with limited $\eta^2 \sim 1\%$ efficiency \cite{sipahigil_integrated_2016, zhang_strongly_2018, sun_cavity-enhanced_2018}. Alternatively, grating structures \cite{hausmann_integrated_2012, rath_grating-assisted_2013} can enable $\eta^2 \sim 10\%$ coupling efficiency by improving mode matching, and can be made broadband using optimized photonic design principles \cite{dory_inverse-designed_2019}.
Nanophotonic structures can also be directly integrated into single-mode fiber networks. This has been done extremely efficiently using adiabatically tapered diamond waveguides coupled to similarly tapered optical fibers using van der Waals forces, yielding efficiencies close to unity \cite{tiecke_efficient_2015, burek_fiber-coupled_2017}.
Currently, this technique requires precise nanopositioning of the adiabatically tapered fiber, which can be challenging and costly to implement in cryogenic conditions where experimental access is limited \cite{evans_photon-mediated_2018, nguyen_integrated_2019}. 
Instead, permanent and efficient integration into large-scale photonic circuits will likely be necessary to scale up nanophotonic cQED experiments. The first steps in this direction have recently been demonstrated using a pick-and-place technique to integrate diamond structures with aluminium nitride photonic circuits, enabling access to group-IV color centers in 72 separate diamond nanophotonic waveguides \cite{wan_large-scale_2020}.

\subsection{Practical considerations for scalability} \label{yield}

A high overall yield of devices usable as coherent spin-photon interfaces is necessary for quantum networking with several nodes \cite{burek_free-standing_2016, nguyen_integrated_2019}.
Beyond deterministic incorporation of defect centers, the monolithic nature of nanofabricated diamond structures creates additional yield-limiting challenges for color center cQED experiments.
Due to imperfections in device fabrication, the resonance wavelength of nanophotonic diamond cavities must be fine-tuned in situ in order to precisely match the low-temperature ZPL emission wavelength of an individual color center.
%
%
Unlike the open Fabry-Perot cavity (Sec.~\ref{fpcav}), in which cavity frequency can be controlled by changing the cavity length, nanophotonic cavities 
cannot be tuned by 
mechanical displacement. Instead, diamond nanophotonic cavities have been tuned by condensing 
gas on the nanostructure at cryogenic temperatures~\cite{mosor_scanning_2005, burek_high_2014, sipahigil_integrated_2016, zhang_strongly_2018}, which increases the local index of refraction, red-shifting the optical mode. This effect can then be reversed selectively and precisely by using laser light to boil away any gas that has already been deposited, leading to a well-controlled net red-shift of individual device frequencies by several percent of the resonance wavelength \cite{nguyen_integrated_2019, evans_photon-mediated_2018}. Although effective, this technique is relatively slow and adds significant experimental overhead in cryogenic engineering.

\begin{figure*}[h!]
	\begin{center}
		\includegraphics[scale=1]{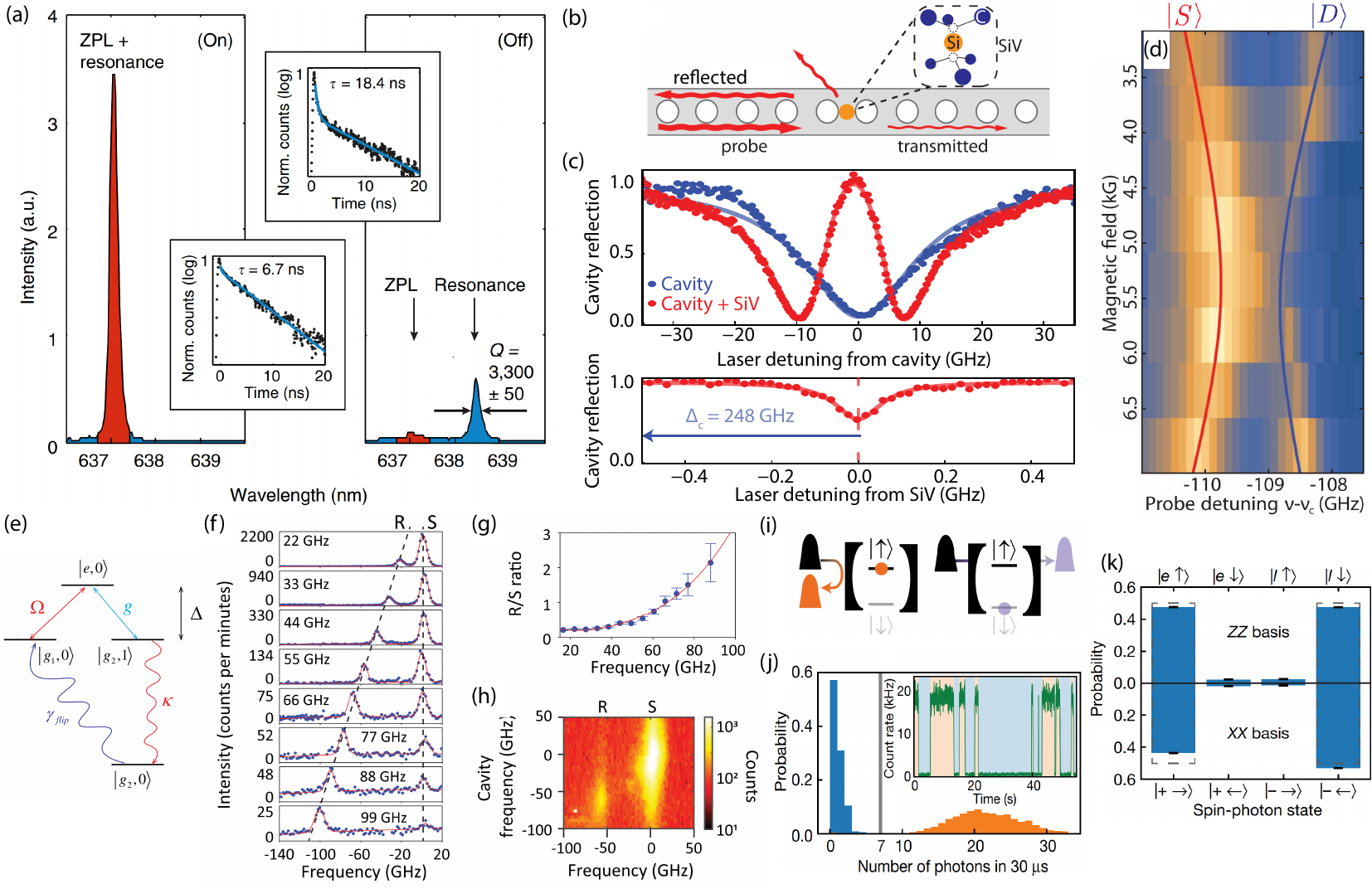}
		\caption{ \label{fig:nexp}
		State of the art diamond nanophotonic cQED experiments.
		(a) Cavity enhancement of NV center ZPL, showing Purcell-reduced excited state lifetime \cite{li_coherent_2015}.
		(b) Schematic for characterizing nanophotonic SiV-cQED parameters by coherent transmission and reflection from SiV-cavity system \cite{sipahigil_integrated_2016}.
		(c) Strongly coupled SiV-cavity system \cite{bhaskar_experimental_2020}. (Top) reflection spectrum of cavity without (blue, $\kappa = 21.6 \pm 1.3$ GHz) and with (red, $g = 8.38 \pm 0.05$ GHz) SiV coupling. (Bottom) Natural linewidth measurement at large cavity detuning ($\gamma = 0.123 \pm 0.01$ GHz), yielding $C = 105 \pm 11$.
		(d) Coherent cavity-photon-mediated interaction between 2 SiV centers tuned into resonance using an external magnetic field \cite{evans_photon-mediated_2018}.
		(e) Level structure and (f) experimental demonstration of cavity-enhanced SiV Raman emission over $\sim 100$ GHz range, with Raman (R) and spontaneous (S) emission components labeled, (g-h) showing transfer of emission spectrum from S to R under suitable cavity detuning \cite{sun_cavity-enhanced_2018}.
		(i) SiV spin-dependent cavity reflection \cite{nguyen_quantum_2019} enables (j) observation of spin quantum jumps and single-shot spin readout with fidelity $F = 0.9998^{+0.0002}_{-0.003}$ in $30~\upmu$s and (k) spin-photon entanglement with state fidelity $F \geq 0.944 \pm 0.008$ \cite{bhaskar_experimental_2020}.
		}
	\end{center}
\end{figure*}

The ability to tune the emission frequency of a color center in-situ is another requirement for integration into quantum networks (Sec.~\ref{defect_section}). Recent nanophotonics experiments have focused on group-IV color centers because they are relatively insensitive to electrical noise from nearby nanostructured surfaces. However, this electric field insensitivity means they also cannot be manipulated by DC Stark shifts. One successful approach has used two-photon Raman transitions to overlap the emission frequencies of two SiV centers that initially differed by several GHz \cite{sipahigil_integrated_2016}.
When employed in combination with cavity coupling, such Raman transitions have enabled widely-tunable, cavity-enhanced single-photon emission spanning the entire inhomogeneous distribution of the SiV center \cite{sun_cavity-enhanced_2018} [Fig.~\ref{fig:nexp}(e-h)].
However, wideband tuning via the Raman approach requires driving fields with substantial strength that may introduce heating and charge instability \cite{sipahigil_integrated_2016, nguyen_integrated_2019}.
%
Alternatively, color center transitions can shift with strain, and electro-mechanical tuning 
has recently enabled deterministic emitter resonance matching and active locking for compensation of spectral diffusion (reduction of $\gamma_d$) inside nanoscale diamond waveguides, two key criteria in building large-scale networks out of solid-state defects \cite{machielse_quantum_2019}.
However, incorporating similar nanomechanical capacitors with free-standing diamond PCCs remains an outstanding challenge.

Incorporation into nanostructures also poses challenges for the spin degrees of freedom of defect centers, where long spin coherence times observed in bulk diamond ~\cite{abobeih_one-second_2018} are reduced by nearby surfaces and impurities introduced by  implantation and fabrication
~\cite{myers_probing_2014, sangtawesin_origins_2019}. Furthermore, state-of-the-art coherence times are achieved using 
lengthy dynamical-decoupling pulse sequences that typically require power delivery on the scale of several watts \cite{lange_universal_2010, pingault_coherent_2017, sukachev_silicon-vacancy_2017, abobeih_one-second_2018, bradley_ten-qubit_2019}. 
Coherent spin control has been accomplished for NV centers inside diamond nanostructures in a helium flow cryostat~\cite{li_coherent_2015}.
However, in the case of SiV centers, 
experiments are typically limited by the milliwatt-scale cooling power of dilution refrigerators at 100 mK temperature \cite{sukachev_silicon-vacancy_2017}. Additionally, lattice strain is required to allow the magnetic dipole transition between spin-states for group-IV color centers \cite{hepp_electronic_2014, sukachev_silicon-vacancy_2017, nguyen_integrated_2019}, reducing the yield of suitable emitters in devices without strain-tuning capabilities.

These challenges have prompted the use of lithographically aligned gold striplines in close proximity to the PCC [Fig.~\ref{fig:nano}(i)]. Such devices have enabled high-fidelity coherent control of an SiV center and nearby $^{13}$C nucleus with coherence times $> 1$ ms and $> 100$ ms respectively at millikelvin temperatures \cite{nguyen_quantum_2019}. Despite this success, residual heating from microwave pulses is believed to limit operational fidelity \cite{bhaskar_experimental_2020}, motivating the development of superconducting microwave striplines on diamond.
As an alternative to microwave manipulation, all-optical spin control using two-photon transitions is possible \cite{yale_all-optical_2013, becker_all-optical_2018}. However, such techniques have yet to yield control fidelities comparable to the microwave approach and have not yet been implemented in nanostructures, which may also be particularly susceptible to heating from optical control fields \cite{meenehan_silicon_2014}.
One final approach is coherent electromechanical driving, which has recently been accomplished using a piezoelectric surface acoustic wave actuator \cite{maity_coherent_2020}.
It is likely that some combination of advances in strain engineering and low-power, on-chip microwave electronics will be required to scale up diamond nanophotonic cQED experiments at cryogenic temperatures.

\subsection{State of the art}

%
The first proof-of-principle experiments almost ten years ago showed Purcell enhancement of NV and SiV ZPL transitions in 
nanophotonic resonators \cite{faraon_resonant_2011, faraon_coupling_2012, hausmann_coupling_2013, riedrich-moller_one-_2012, riedrich-moller_deterministic_2014, li_coherent_2015} [see Fig. \ref{fig:nexp}(a) for an example].
However, these experiments were limited to the $C < 1$ regime due to emitter dephasing inside heavily fabricated structures, which remains an outstanding challenge for NV centers \cite{faraon_coupling_2012}.
Instead, the use of environmentally insensitive SiV centers has enabled pioneering experiments in the $C \gtrsim 1$ regime, which was experimentally verified using the coherent extinction of resonant transmission through the cavity [Fig. \ref{fig:nexp}(b)], as well as radiative broadening of the SiV optical transition \cite{sipahigil_integrated_2016, zhang_strongly_2018}.

Technical improvements, including more accurate emitter positioning using lithographically aligned masks \cite{nguyen_integrated_2019}, more reliable angled-etch procedures using ion-beam milling \cite{atikian_freestanding_2017}, and operation at millikelvin temperatures to eliminate residual thermal decay and decoherence effects \cite{sipahigil_integrated_2016, sukachev_silicon-vacancy_2017} have now enabled nanophotonic SiV cQED devices in the $C \gg 1$ regime [Fig.~\ref{fig:nexp}(c)], \cite{bhaskar_experimental_2020}.
This has allowed the first experimental observation of photon-mediated interactions between two emitters inside of a cavity using SiV centers \cite{evans_photon-mediated_2018} [Fig.~\ref{fig:nexp}(d)], offering a potential pathway towards deterministic quantum logic gates between color centers similar to those employed in superconducting microwave quantum processors \cite{majer_coupling_2007}. Translation of these techniques to the optical domain will require further improvements in device cooperativity \cite{borregaard_quantum_2019} along with implementation of optimized heralded schemes to overcome cavity loss \cite{borregaard_heralded_2015}.

Coherent spin-photon interfaces have been achieved by combining these high cooperativity devices with microwave spin control of SiV centers and nearby nuclear spins \cite{sukachev_silicon-vacancy_2017, nguyen_quantum_2019}. Additional technical advances, such as vector magnetic field control and precise optimization of SiV-cavity detuning, have given access to SiV spin states with high cavity reflection contrast and cycling transitions [Fig.~\ref{fig:nexp}(i)] \cite{nguyen_integrated_2019}. These elements allow for extremely high-fidelity single-shot spin readout [Fig.~\ref{fig:nexp}(j)] and spin-photon entanglement [Fig.~\ref{fig:nexp}(k)] \cite{bhaskar_experimental_2020}. 

The current generation of diamond nanophotonic cavities are among the state of the art across all experimental platforms in several key quantum networking criteria: atom-photon cooperativity ($C > 100$) \cite{bhaskar_experimental_2020}, spin coherence times ($T_2^{\mathrm{SiV}} > 1$ ms, $T_2^{^{13}C} > 0.2$ s) \cite{nguyen_quantum_2019}, spin readout and spin-photon entanglement fidelity ($F_r = 0.9998^{+0.0002}_{-0.0003}, F_e \geq 0.944 \pm 0.008$) \cite{bhaskar_experimental_2020}, and emitter tunability ($\sim 100$ GHz) \cite{sun_cavity-enhanced_2018, machielse_quantum_2019}.
These developments have very recently culminated in the first experimental demonstration of a memory-enhanced quantum communication protocol \cite{bhaskar_experimental_2020}.
Diamond nanophotonic cavities are now being used for implementation of novel protocols in quantum optics and information science reaching beyond the diamond photonics community, signaling the maturity of the field and its potential to play a prominent role 
in construction of quantum technologies.

\subsection{Outlook} 

Despite immense progress 
in the field, diamond nanophotonics has yet to approach the peak of its potential. A number of outstanding challenges remain to be tackled.
First, truly scalable fabrication will require development of wafer-scale quantities of thin-film diamond \cite{butler_thin_1993, jiang_heteroepitaxial_1993, schreck_diamondirsrtio3_1999, teraji_chemical_2006, rath_waferscale_2013, habibpiracha_scalable_2016, nelz_toward_2019} for rapid and reliable processing into photonic crystal cavities.
Second, in the short term, current techniques for underetching structures in diamond can still be improved, as quality factors of nanoscale PCCs are still limited by scattering losses introduced by imperfections in device fabrication \cite{burek_high_2014, mitchell_realizing_2019}. 
%
%
Finally, integration of diamond PCCs with more sophisticated nanophotonic circuits will be required to truly leverage the advantages of nanophotonics for large-scale systems~\cite{mouradian_scalable_2015, wan_large-scale_2020}. Efficient incorporation of detectors \cite{atikian_superconducting_2014, rath_superconducting_2015}, routing elements such as beamsplitters and switches \cite{qiang_large-scale_2018, papon_nanomechanical_2019}, control and tuning electronics \cite{machielse_quantum_2019, nguyen_integrated_2019}, and frequency conversion to telecommunications wavelengths \cite{dreau_quantum_2018, maring_quantum_2018} into large-scale fiber networks \cite{burek_fiber-coupled_2017} remains a long-term goal for diamond nanophotonic cQED systems.

\section{Conclusion}
This review has considered two approaches for cavity coupling of diamond defects with complementary strengths. Fabry-Perot-style cavities offer compatibility with bulk-like defects, making them suitable for use with NV centers in currently available materials~\cite{ruf_optically_2019}. Moreover, their very high quality factors $>10^6$ allow linewidths smaller than the fine structure splittings of the NV center, enabling the spin-selective cavity enhancement employed in proposed protocols for qubit readout and quantum networks~\cite{nemoto_photonic_2014, nemoto_photonic_2016, hanks_high-fidelity_2017}. Finally, cavity mirrors formed on the tips of optical fibers couple directly to propagating fiber modes, simplifying outcoupling. Diamond nanophotonic approaches, on the other hand, achieve nanoscopic mode volumes while maintaining good quality factors, leading to the strongest Purcell enhancements. While NV centers broaden problematically in such structures, the SiV retains good optical coherence, and newer defects may prove similarly insensitive to nanofabrication~\cite{bradac_quantum_2019}. The different capabilities of the two platforms make them suited to complementary goals in quantum information science. 


In the longer term, both cavity platforms will require improvements and additional capabilities to realize practical applications of quantum networks. Many technical challenges require further attention, including inhomogeneous broadening on emitter optical transitions and intra-cavity control over spins. Beyond improving the properties of the cQED platform itself, other advances will be needed to incorporate them into quantum network applications. Since diamond defects emit primarily in the visible and NIR, conversion to telecom wavelengths will be necessary for long-distance networks. Already, 17\% conversion efficiency has been achieved for the NV center~\cite{dreau_quantum_2018}, and even employed to demonstrate spin-photon entanglement at 1588 nm~\cite{tchebotareva_entanglement_2019}. A final challenge is scaling up: while initial quantum repeater demonstrations could be achieved with single cavity-coupled defects (with a few auxiliary nuclear spin qubits), more advanced applications -- for example those employing error correction -- will likely require an increased number of qubits per node. While some on-chip integration may be possible for arrays of open microcavities~\cite{derntl_arrays_2014, trupke_large-scale_2016}, nanofabricated approaches offer a clear opportunity to combine cavity-coupled qubits with integrated photonics routing structures~\cite{hausmann_integrated_2012, faraon_quantum_2013, mouradian_scalable_2015} for large-scale on-chip applications~\cite{elshaari_hybrid_2020, kim_hybrid_2020, wan_large-scale_2020}. 

In conclusion, while challenges still lie ahead, recent advances in cavity coupling of diamond defects have demonstrated the feasibility of realizing a coherent interface between single photons and long-lived solid-state spins. Such a platform places the field on the cusp of finally realizing many of the ideas that first inspired interest in diamond defects, and it may someday form the fundamental building block for practical quantum network applications. 

\noindent{\bf Acknowledgements.} 
The authors thank Yannik Fontana for providing the defect diagrams used in Fig. 2, Christian Nguyen for providing the SiV spectrum used in Fig. 2, Bartholomeus Machielse for giving comments and providing the false-color SEM in Fig. 5,  and Paul Barclay, Stefan Bogdanovic, and Daniel Riedel for providing insightful commentary on the manuscript. 

\noindent{\bf Funding.} M. K. Bhaskar acknowledges support from an NDSEG fellowship. L. Childress is a CIFAR fellow in the Quantum Information Science program, and acknowledges support from Canada Research Chairs projects 229003 and 231949, Canada Foundation for Innovation projects 229003 and 33488, Fonds de Recherche - Nature et Technologies FQRNT PR-253399, National Sciences and Engineering Research Council of Canada NSERC RGPIN 435554-13 and RTI-2016-00089 and RGPIN-2020-04095, and  l’Institut Transdisciplinaire d’Information Quantique (INTRIQ).

\noindent{\bf Disclosures.} The authors declare no conflicts of interest.

\bibliography{optica_20200524}

\end{document}